\documentclass[a4paper,12pt]{article}

\title{\LARGE\bf A simple asymmetric evolving random network}
\date{}
\author{}
\newcommand{\bin}[2]{{ #1 \choose #2 }}
\newcommand{\mona}{\alpha}
\newcommand{\infb}{b}
\newcommand{\supb}{\beta}
\newcommand{\mont}{\tau}
\newcommand{\vev}[1]{\langle #1 \rangle}
\usepackage[final]{graphicx}
\begin{document}
\maketitle

\vspace{-1.2cm}

\centerline{\large Michel Bauer\footnote[1]{Email:
    bauer@spht.saclay.cea.fr} and Denis Bernard\footnote[2]{Member of
    the CNRS; email: dbernard@spht.saclay.cea.fr}} 

\vspace{.3cm}

\centerline{\large Service de Physique Th\'eorique de
  Saclay\footnote[3]{\it Laboratoire de la Direction des Sciences de
    la Mati\`ere du Commisariat \`a l'Energie Atomique, URA2306 du CNRS}}

\vspace{.3cm}

\centerline{\large CE Saclay, 91191 Gif sur Yvette, France}

\vspace{.3cm}

\begin{abstract} We introduce a new oriented evolving graph model
  inspired by biological networks. A node is added at each time step
  and is connected to the rest of the graph by random oriented edges
  emerging from older nodes. This leads to a statistical asymmetry
  between incoming and outgoing edges. We show that the model exhibits
  a percolation transition and discuss its universality. Below the
  threshold, the distribution of component sizes decreases
  algebraically with a continuously varying exponent depending on the
  average connectivity.  We prove that the transition is of infinite
  order by deriving the exact asymptotic formula for the size of the
  giant component close to the threshold. We also present a thorough
  analysis of aging properties. We compute local-in-time
  profiles for the components of finite size and for the giant
  component, showing in particular that the giant component is always
  dense among the oldest nodes but invades only an exponentially small
  fraction of the young nodes close to the threshold.
\end{abstract}


\section{Motivations and results.}

Evolving random graphs have recently attracted attention, see
e.g. refs \cite{barab,doro1,new} and references therein. 
This interest is mainly motivated by
concrete problems related to the structure of 
communication or biological networks. 
Experimental data are now available in many contexts \cite{barab,kepes}. 

In these examples, the asymmetry and the evolving nature of the
networks are likely to be important 
ingredients for deciphering their statistical properties. 
It is however far from obvious to find solvable cases that would
possibly account for some relevant features of, say, the regulating
network of a genome. Although biology has strongly influenced
our interest in evolving networks, the model we solve is not based on
realistic biological facts but it nevertheless incorporates asymmetry
and chronological order. 
Understanding such simple evolving graphs may help understanding 
biological networks, at least by comparison and opposition.

We were initially motivated by the study of the yeast genetic
regulatory network presented in ref.\cite{kepes}. The 
authors studied in and out degree distributions and discovered a
strong asymmetry: a single gene may participate to the regulation of
many other genes -- the law for out-degrees seems to be large --,
but each genes is only regulated by a few other genes -- the law for
in-degrees seems to have finite moments. This is why we consider oriented
evolving random graphs in the sequel. A biological interpretation for
the asymmetry is that the few promoter-repressor sites for each gene
bind only to specific proteins, but that along the genome many
promoter-repressor sites are homologous. However, this does not
predict the precise laws. An understanding of the same features
from a purely probabilistic viewpoint would be desirable as well.  

The recent experimental studies dealt with global statistical
properties of evolving graphs, i.e. when the evolving network is 
observed at some fixed time with the ages of different vertices and
edges not taken into account. There are simple experimental reasons for that~:
to keep track of the ages would in many cases dramatically reduce 
the statistics, and in other cases this information is even not available. 
Our second motivation is a better understanding of the local-in-time 
statistical properties of evolving networks. This helps dating 
or assigning likely ages to different structures of the networks. 
As we shall later see, the global analysis, which is like a
time average, gives a distorted view of the real structure of the
networks. We shall present a detailed analysis of local-in-time
features in our model.  

\vspace{.3cm}

The model we study is the natural evolving cousin of the famous
Erd\"os-Renyi random graphs \cite{erdos}. Starting from a single
vertex at time $1$, a new vertex is created at each time step -- so
that at time $t$, the size of the system, i.e. the number of vertices,
is $t$ --, and new oriented edges are created with specified
probabilistic rules. A tunable parameter $\mona$ ranging from $0$ to
$\infty$ describes asymptotically the average number of incoming edges
on a vertex. Precise definitions are given in the next section.

Our main results are the following~:

From very simple rules, we see an asymmetry emerging. The global in
and out degree distributions are different. We also compute the local
profiles of in and out degree distributions, and comment on the
differences.

We make a detailed global analysis for
the structure and sizes of the connected components. We use generating
function methods to write down a differential equation that implies
recursion relations for the distribution of component sizes, see 
eqs.(\ref{Cdiff},\ref{Crecur}). 

A salient global feature of the model is a percolation phase
transition at a critical value of the average connectivity. Below this
value, no single component contains a finite fraction of the sites in
the thermodynamic limit, i.e. in the large $t$ limit.  However, a
slightly unusual situation occurs in that below the transition the
system contains components whose sizes scale like a power of the total size
of the graph, see eq.(\ref{eq:grosclu}).  Correspondingly, the
probability distribution for component sizes has an algebraic queue,
see eq.(\ref{asympk}), and its number of finite moments jumps at
specific values of the average connectivity. Above the transition,
this probability distribution becomes defective, but its decrease is
exponential, see eq.(\ref{Pklarge}). The transition is continuous.
Close to the threshold, the fraction of sites in the giant component
-- the percolation cluster -- has an essential singularity, see
eq.(\ref{eq:pof}).  We argue that this result is universal, with the
meaning used in the study of critical phenomena. The essential
singularity at the percolation threshold had already been observed
numerically by \cite{new} in a different model which we show to be in
the same universality class as ours for the percolation transition,
and computed analytically for another class of models in \cite{doro2}.

We then turn to the study of local-in-time profiles of connected
components. Guided by a direct enumeration based on tree
combinatorics, we show that they satisfy recursion relations, and we
give the first few profiles (isolated vertices, pairs, triples)
explicitly. The profile of the giant component is given by a
differential equation, from which we extract the singularity in the
far past and the critical singularity in the present -- see
eqs(\ref{eq:rho_0},\ref{eq:rho_1}). In particular the giant component
invades all the time slices of the graph above the transition. One
strange feature of profiles, which would deserve a good explanation,
is that in several instances the formal parameter involved in
generating functions for global quantities is simply traded for the
relative age to obtain interesting local-in-time observables, see
eqs.(\ref{eq:young},\ref{Ddiff}).

We have compared our analytical results with numerical simulation whenever
possible. 

\vspace{.3cm}

While polishing this paper, we became aware of \cite{kim}, 
whose goals overlap partly with ours. When they can be compared,
the results agree. 

\section{The model.}

We construct evolving random graphs with the following rules:\\
(i) We consider a triangular array of independent random variables
$\ell_{i,j}$, $1\leq i<j$, where $\ell_{i,j}$ takes value $1$ with
probability
$p_j\in[0,1]$ and value $0$ with probability $q_j\equiv 1-p_j$.\\
(ii) We start from the graph made of single vertex at initial time
$t=1$. At time $t$, $t\geq 2$, a vertex with label $t$ is added
together with the {\it directed} edges $[j\to t]$ for which
$\ell_{j,t}=1$.
We shall often take the viewpoint that the (biased) coin tossings
defining  $\ell_{i,t}$  are done at time $t$.

We shall assume that $p_t\simeq \mona  /t$ at large time $t$,
with $\mona $ a parameter which we shall identify as half the average
connectivity. 
This choice ensures the convergence of various distributions to
stationary measures, most of them being independent of the
precise values of the early probabilities.

By construction  all edges arriving at a given vertex are 
simultaneously created at the instant of creation of this vertex.
As a consequence, these graphs are not only oriented but
chronologically oriented -- this is unrealistic from the biological
viewpoint.

\section{Edge distributions.}

In this section we give the incoming and outgoing edge distributions.\\
Let  $\ell_j^-(t)$ be the number of incoming edges at the vertex $j$,
and $\ell_j^+(t)$ be the number of outgoing edges at this vertex at
time $t$.\\
Let $v_k^-(t)$ be the number of vertices with $k$ incoming edges,
and $v_k^+(t)$ be the number of vertices with $k$ outgoing edges at
time $t$.

We may look either at the edge distributions at a given vertex,
or we may look at the edge distributions defined by gathering
averaged histograms over whole graphs.
The former are specified by their generating functions,
$$
\vev{ z^{\ell_j^\pm(t)}}
$$
where $\vev{\cdots}$ denotes expectation value.
It may depend on the specified vertex labeled by $j$.
The latter is defined by the generating functions,
\begin{eqnarray}
{\mathcal V}^\pm_t(z) \equiv \frac{1}{t} \sum_{0\leq k \leq t} 
\vev{v_k^\pm(t)}\,
z^k = \frac{1}{t} \sum_{1\leq j\leq t} \vev{z^{\ell_j^\pm(t)}} 
\label{histo}
\end{eqnarray}
We remark that this global histogram distribution is the
average of the local-in-time quantity $\vev{z^{\ell_j^\pm(t)}}$. Since
at time $t$ the total number of vertices is $t$, ${\mathcal
  V}^\pm_t(z)$ is properly normalized, ${\mathcal V}^\pm_t(1)=1$, to
define an averaged probability distribution function, independent of
the vertices, for the incoming or outgoing edge variables $\ell^\pm$:
$$
 \sum_k z^k\, {\rm Prob}_{(\ell^\pm=k)} \equiv {\mathcal V}_t^\pm(z)
$$

{\it Incoming vertices.} The number of incoming edges
$\ell_j^-(t)=\sum_i\, \ell_{i,j}$ at vertex $j\leq t$ asymptotically possesses
a Poisson distribution since
$$
\vev{ z^{\ell_j^-(t)}} =
    \prod_{i<j}[q_j+zp_j]\simeq_{j\to\infty}\exp(\mona(z-1))
$$
The convergence of this distribution justifies our choice of
asymptotic probabilities $p_j\simeq \mona /j$.
Only the vertices whose ages $j$ scale with the age of the graph,
i.e. with $j/t=\sigma$ fixed, $0\leq \sigma\leq 1$, give non trivial
contributions at large time to the averaged histogram (\ref{histo}) and 
\begin{eqnarray}
{\mathcal V}^-_t(z) \simeq_{t\to\infty} \exp(\mona (z-1))
\label{v-}
\end{eqnarray}
This expression may also be retrieved by looking at the evolution
equation of ${\mathcal V}^-_t(z)$. Indeed, consider adding the new
vertex at time $t$. Since the edges are oriented from older to younger
vertices \footnote{A vertex is older than another if it appeared
  before in the evolution, i.e. if it corresponds to a smaller value
  of $\sigma$.}, we have $t{\mathcal V}^-_t(z)=(t-1){\mathcal
  V}^-_{t-1}(z) +\vev{z^{\ell^-_t(t)}}$ from the second definition in
eq.(\ref{histo}).  This is equivalent to
\begin{eqnarray}
t{\mathcal V}^-_t(z)=(t-1){\mathcal V}^-_{t-1}(z)+
(q_t+zp_t)^{t-1}\nonumber 
\end{eqnarray}
As $(q_t+zp_t)^{t-1}\simeq e^{\mona (z-1)}$ at large time, the
stationary limit is given by eq.(\ref{v-}).  This yields a Poissonian
distribution with probabilities
\begin{eqnarray}
{\rm Prob}_{(\ell^-=k)}= e^{-\mona}\,\frac{\mona^k}{k!}
\label{prob-}
\end{eqnarray}

{\it Outgoing vertices.} At a given vertex $j\leq t$, with $j/t=\sigma$
fixed, the number of outgoing edges $\ell_j^+(t)=\sum_{i\leq t} \ell_{j,i}$ at
vertex $j$ also have a Poisson distribution at large time,
$$
\vev{ z^{\ell_{\sigma t}^+(t)}} =
    \prod_{\sigma t=j<i\leq t}[q_i+zp_i]
\simeq_{t\to\infty} \exp(-\mona\log\sigma(z-1))
$$
but with a parameter $\mona \log(1/\sigma)$ depending on the age of the
vertex. Approximating at large time the sum over $j$ in
eq.(\ref{histo}) by an integral over $\sigma$ gives the histogram
distribution:  
\begin{eqnarray}
{\mathcal V}^+_t(z) \simeq_{t\to\infty} \int^1_0d\sigma\,
 \vev{ z^{\ell_{\sigma t}^+}(t)} = \frac{1}{1+ \mona (1-z)}
\label{v+}
\end{eqnarray}
As for incoming vertices, this formula follows from the evolution
equation for ${\mathcal V}^+_t(z)$. Indeed, since the numbers of
outgoing edges $\ell^+_j(t)$ from vertex $j$ at time $t$ and $t-1$ differ
by $\ell_{j,t}$ we have $\vev{z^{\ell^+_j(t)}}=
\vev{z^{\ell^+_j(t-1)}}\, \vev{z^{\ell^+_{j,t}}}$. From definition
(\ref{histo}) this gives
\begin{eqnarray}
t{\mathcal V}^-_t(z)=1+(t-1){\mathcal V}^-_{t-1}(z)\,(q_t+zp_t) \nonumber
\end{eqnarray}
where the first term is the contribution of the newly added vertex at
time $t$. The stationary limit is given by eq.(\ref{v+}).
This is a geometric  distribution, slightly larger than the Poisson
distribution, with probabilities
\begin{eqnarray}
{\rm Prob}_{(\ell^+=k)} = \frac{\mona ^k}{(1+\mona )^{k+1}}
\label{prob+}
\end{eqnarray}

{\it Mixed distribution}. Let $v_{k_+,k_-}(t)$ be the number of vertices
with $k_+$ outgoing and $k_-$ incoming edges at time $t$. As in eq.(\ref{histo}),
the generating function for the mixed histogram distribution is defined by
\begin{eqnarray}
{\mathcal V}_t(z_+,z_-)\equiv \frac{1}{t}\sum_{k_+,k_-}
\vev{v_{k_+,k_-}(t)}\,z_+^{k_+}z_-^{k_-}
= \frac{1}{t}\sum_{1\leq j\leq t}\vev{z_+^{\ell_j^+(t)}z_-^{\ell_j^-(t)}}
\nonumber
\end{eqnarray}
By construction the outgoing and incoming edges variables $\ell_j^\pm(t)$
are statistically independent for $j$ fixed, so that the last
expectation values factorize. As above we may derive an evolution
equation by evaluating the contribution of the newly added vertex at
time $t$. This yields:
\begin{eqnarray}
t{\mathcal V}_t(z_+,z_-)= (q_t+z_-p_t)^{t-1} + (t-1){\mathcal 
V}_{t-1}(z_+,z_-)\, (q_t+z_+p_t) 
\nonumber
\end{eqnarray}
Its stationary limit is factorized:
\begin{eqnarray}
{\mathcal V}(z_+,z_-) =\frac{e^{\mona (z_--1)}}{1+\mona (1-z_+)}  \nonumber
\end{eqnarray}
Outgoing and incoming edges are statistically
independent at large time.

\section{Cluster distributions.}
\label{sec:clusters}

In this section, we present the main relations governing the
probability distributions of connected components of the graphs. Two
vertices belong to the same connected component if they can be joined
by a path made of edges, without any reference to orientation. This
definition ensures that the property of being in  the same connected
component is an equivalence relation, but does not ensure that two
points in a connected component can be joined by an oriented path. 

To partly avoid repetitions, the term \textit{cluster} is used as a
synonymous for \textit{connected component} in the sequel.

Intuitively, the fact that the network is fragmented can be understood
as follows : when a vertex $t_0$ is created, it has a finite
probability to be isolated, and the probability that none of the
vertices $t_0+1,\cdots,t$ connects to vertex $t_0$ is $q_{t_0+1}\cdots
q_t$ which scales as $(t_0/t)^\mona$. This quantity remains finite as
long as $t_0/t$ does.  This argument shows that there are isolated
vertices in the system. A small extension of the argument shows that
there are also finite components and that young
vertices are more likely to be in small components than old ones. This
will be made more rigourous in the study of profiles, see
section \ref{sec:chropro}.

Let $N_k(t)$ be the number of connected components with $k$ vertices at
time $t$ and let $N_t(z)$ be the generating function,
$$
N_t(z) = \sum_{k\geq 1} N_k(t)\, z^k
$$
By definition, $\sum_kN_k(t)$ is the number of components and $\sum_k
N_k(t) k$ the total number of vertices, $\sum_k N_k(t) k=t$ at any
finite time.

Let us write an evolution equation for $N_t(z)$. 
At time $t+1$, we add the vertex with label $t+1$ which may then be
connected to $n_k(t)$ connected components of size $k$. This creates
a new component of size $1+\sum_k n_k(t)k$, but also removes $n_k(t)$
components of size $k$. Thus, at time $t+1$ we have:
\begin{eqnarray}
N_k(t+1) = N_k(t) -n_k(t)+ \delta_{k;1+\sum_pn_p(t)p}
\label{Nt}
\end{eqnarray}
with $\delta_{j;k}$ the Kronecker symbol. Alternatively,
\begin{eqnarray}
N_{t+1}(z) = N_t(z) - \sum_{k\geq 1} n_k(t)z^k + z^{1+\sum_k n_k(t)k}
\label{Ntbis}
\end{eqnarray}

As is apparent from this formulation, the transition probability 
from a given $N_t(z)$ to a given $N_{t+1}(z)$ can be given in closed
form. To be precise, the admissible $N_t(z)$'s (describing the
accessible distributions of components at time $t$) are polynomials
with integral non-negative coefficients, whose derivative at $z=1$ have
value $t$. Now suppose $N_t(z)$ and $N_{t+1}(z)$ are admissible. If
the difference $N_{t+1}(z) - N_t(z)$ cannot be written as $- \sum_{k\geq 1}
n_k(t)z^k + z^{1+\sum_k n_k(t)k}$ for some set of nonnegative integers
$n_k(t)$, the transition is forbidden. If it can, then the $n_k(t)$'s
are uniquely defined and the transition probability is 
\[
{\rm Prob}(N_t(z) \rightarrow N_{t+1}(z))=\prod_k \bin{N_k(t)}{n_k(t)}
q_{t+1}^{k(N_k(t)-n_k(t))}(1-q_{t+1}^k)^{n_k(t)}
\]
The meaning of this equation is simple. At time $t+1$, the new vertex
is added, and for each of the former $t$ points a (biased) coin is
tossed to decide the value of the edge variables
$\ell_{j,t+1}$. The tossings are independent with the same law, so
the probability that the new point does not attach to a given
component of size $k$ is $q_{t+1}^k$, and distinct components are
independent. Hence for each $k$ one makes
$N_k(t)$ independent Bernoulli trials with failure probability
$q_{t+1}^k$, and the transition from $N_t(z) \rightarrow  N_{t+1}(z)$ requires
exactly $n_k(t)$ successes.
This shows that the graph evolution is a (time inhomogeneous) Markov 
process on the space of components distributions, a fact that we shall
use for the purpose of numerical simulations. 

This explicit representation of the transition probability could be
used to average equation (\ref{Ntbis}). Alternatively, one can
represent the number $n_k(t)$ of components of size $k$ which
are connected to the new vertex in terms of the edge variables
$\ell_{i,j}$ as
$$
n_k(t) = \sum_{[k]=1}^{N_k(t)}[1 - \prod_{j\in[k]}(1-\ell_{j,t+1})]
$$
where $[k]$ runs over connected components of size $k$. Since the
edge variables $\ell_{j,t+1}$ are statistically independent of the
earlier edge variables, $\ell_{j,k}$ with $k\leq t$, and therefore
also independent of the $N_k(t)$'s, we have for any $w$,
\begin{eqnarray}
\vev{w^{n_k(t)}}= \vev{[q_{t+1}^k+(1-q_{t+1}^k)w]^{N_k(t)}}
\label{ndist}
\end{eqnarray}
In particular,
\begin{eqnarray}
\vev{n_k(t)}= (1-q_{t+1}^k)\, \vev{N_k(t)}
\label{nbis}
\end{eqnarray}
We can now take the average value of eq.(\ref{Ntbis}) to get
\begin{eqnarray}
\vev{N_{t+1}(z)} & = & \vev{N_{t}(zq_{t+1})}
+ z \vev{\prod_{k\geq 1}[q_{t+1}^k+(1-q_{t+1}^k)z^k]^{N_k(t)}}
\nonumber \\
& = & \vev{N_{t}(zq_{t+1})}
+ z q_{t+1}^t \vev{\prod_{k\geq 1}[1+(q_{t+1}^{-k}-1)z^k]^{N_k(t)}}
\label{Nmoyen}
\end{eqnarray}

Assume, but this will be justified later using the asymptotic
behavior of the $p_j$'s, that at large time and for fixed component size
$N_k(t)/t$ is self-averaging, meaning that 
$$
N_k(t)/t \simeq  C_k + o(1),\quad {\rm as}\ t\to\infty
$$
with $C_k$ equals to its averaged value with probability one and
with the remaining other $o(1)$ terms random.  This in particular
implies that for any finite size $k$ the number of connected
components of size $k$ scales thermodynamically with the graph size.
If both sides of eq.(\ref{Nmoyen}) are expended in powers of $z$, a
given degree involves only components of bounded ($t$-independent)
size.  So, order by order in $z$, self-averaging applies and we
conclude that taking $1-q_t \sim \mona /t$, the following is an
accurate approximation for the last term in eq.(\ref{Nmoyen}) at
large times:
$$
 q_{t+1}^t\prod_{k\geq 1}[1+(q_{t+1}^{-k}-1)z^k]^{N_k(t)}
\simeq_{t\to \infty} \exp(-\mona  +\mona \sum_{k\geq 1}kz^kC_k).
$$
The averaged evolution equation (\ref{Nmoyen}) then gives a
deterministic differential equation
\begin{eqnarray}
0=- C(z) - \mona  z\partial_z C(z) + z \exp(-\mona  + \mona
z\partial_z C(z)) 
\label{Cdiff}
\end{eqnarray} 
for the generating function 
\[
C(z)\equiv \sum_{k\geq 1}z^kC_k. 
\]
The function 
\[
z\partial_z C(z) = \sum_{k\geq 1} z^kP_k, \quad P_k\equiv kC_k
\]
has a direct probabilistic interpretation. Indeed, $P_k$ is the
fraction of points in clusters of size $k$, or equivalently the
probability that a randomly chosen vertex belongs to a connected
component of size $k$.  

By construction $C(z)$ is a Taylor series with
positive coefficients. The series $\sum_k kC_k=\sum_k P_k$ is
convergent because it counts the fraction of points in finite clusters
which is $\leq 1$. As a consequence the radius of convergence of
$C(z)$ is at least $1$: if we denote by $R(\mona)$ the radius of
convergence $C(z)$, we know that $R(\mona)\geq 1$ for $0 \leq \mona <
+\infty$.

We show in Appendix A that analogous methods can be used to describe
the connected components of the  Erd\"os-Renyi random graph model. 

Most of our subsequent analysis will be based on (\ref{Cdiff}).
It recursively determines the $C_k$'s. The first few are:
\begin{eqnarray}
C_1&=& \frac{e^{-\mona }}{(\mona +1)},\
C_2=\frac{\mona  e^{-2\mona }}{(\mona +1)(2\mona +1)},\label{lesC}\\
C_3&=&\frac{\mona ^2(6\mona +5)e^{-3\mona }}{2(\mona +1)^2(2\mona
  +1)(3\mona +1)},\ \cdots\nonumber 
\end{eqnarray}
More generally, the combinations $C_ke^{k\mona }$ are rational
functions in $\mona $.  They are more efficiently computed using a
second order differential equation which follows from derivation of
the logarithm of eq.(\ref{Cdiff}):
\begin{eqnarray}
z\partial_z(C+\mona  z\partial_zC)=(C+\mona  z\partial_zC)(1
+ \mona(z\partial_z)^2C) \label{Cd2} 
\end{eqnarray}
This leads to
\begin{eqnarray}
(n-1)(1+n\mona )C_n=\sum_{k+l=n} k^2(1+\mona l)C_kC_l   
\label{Crecur}
\end{eqnarray}
A general feature of the recursion relations is that the rational
function $C_ke^{k\mona }$ has no poles except possibly at the
points $-1,-1/2,\cdots,-1/k$.

Eq.(\ref{Nmoyen}) is not a closed formula for the
probability distributions of the number of connected components.
But as shown in Appendix B, one may improve slightly the above
argument to obtain a closed system (see eq.\ref{mast}) which can be
used to prove systematically that the variables $C_k$ 
are self-averaging, a task we perform for $C_1$ and $C_2$ in Appendix
B. The proof becomes more and more tedious for large $k$. But we are
confident that the self averaging property is general because of the 
close relationship with the Erd\"os-Renyi model presented
in Appendix A.

\section{Percolation transition.}

Up to now, our analysis of connected components has always
concentrated on finite components: the thermodynamic limit
$t\rightarrow \infty$ has been
taken while keeping $k$, the component size, arbitrary but fixed. 
We have argued that in this regime, the number of components of size
$k$ is proportional to $t$. This has lead to a satisfactory
description of $C_k$. Our arguments did not use
any hypothesis on whether or not  only finite components play a role
in the thermodynamic limit. However, as we observed already, the sum
$\sum_k P_k$ measures 
exactly the fraction of the sites that are either isolated or in
components of size $2$ or $3$ or $\cdots$. To rephrase it more
vaguely, $\sum_k P_k$ counts the fraction of vertices that belong 
to finite connected components. If only clusters of finite size 
contribute to the thermodynamic limit, then $\sum_k P_k=1$. Else, by
standard physical arguments, a single giant component of size $\sim
t(1-\sum_k P_k)$ accounts for the deficit. The giant component is
also called the percolating cluster. Its relative size, which
we denote by $P_\infty$, is
\begin{eqnarray}
P_\infty\equiv 1 - \sum_{k\geq 1} P_k = 1 - \partial_zC(1)
\label{dinfty}
\end{eqnarray}

\begin{figure}[htbp]
  \begin{center}
\begin{minipage}[t]{0.47\textwidth}\vspace{0pt}
\centering \includegraphics[width=\textwidth]{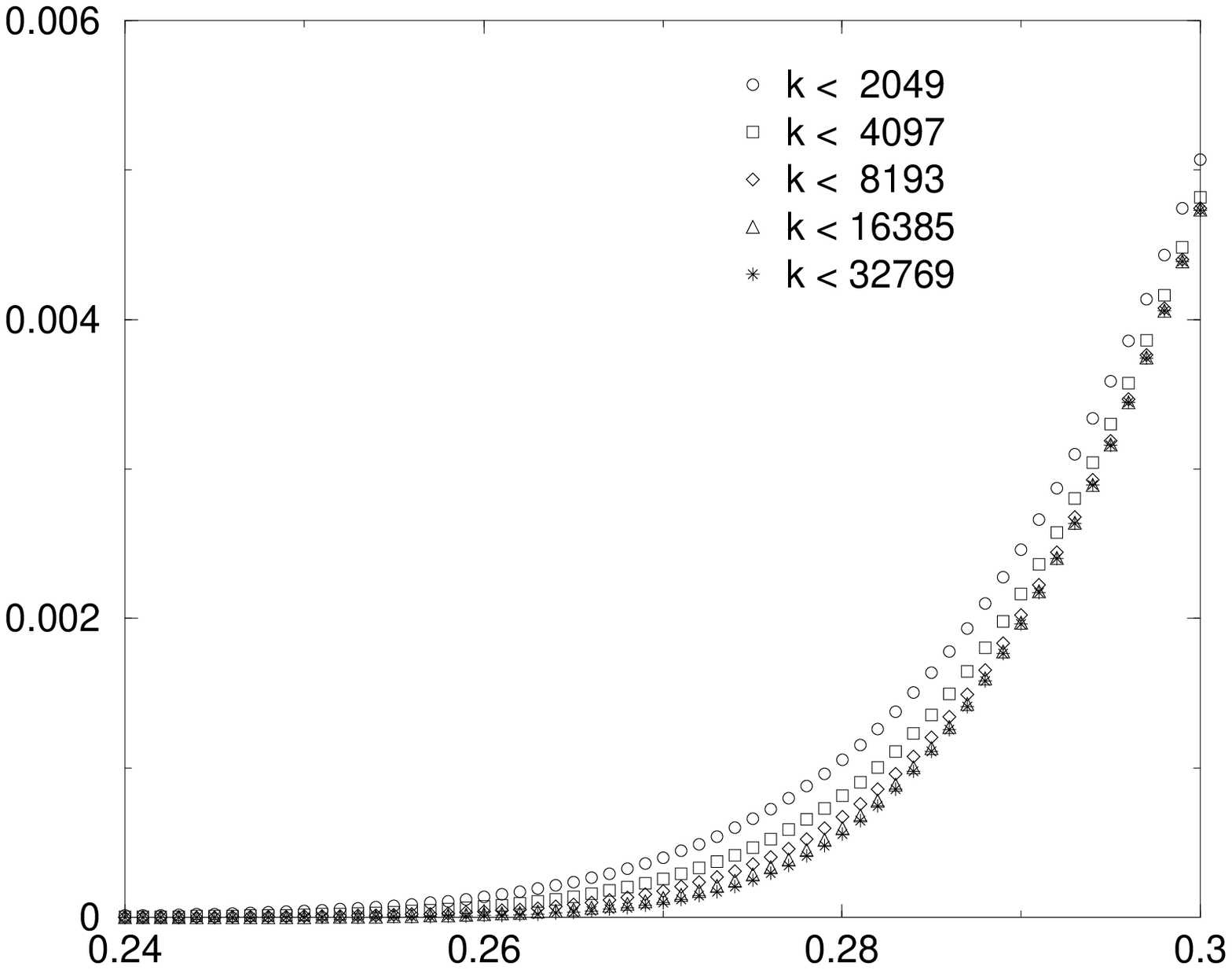}
\end{minipage} \hfill
\begin{minipage}[t]{0.45\textwidth}\vspace{0pt}
\centering \includegraphics[width=\textwidth]{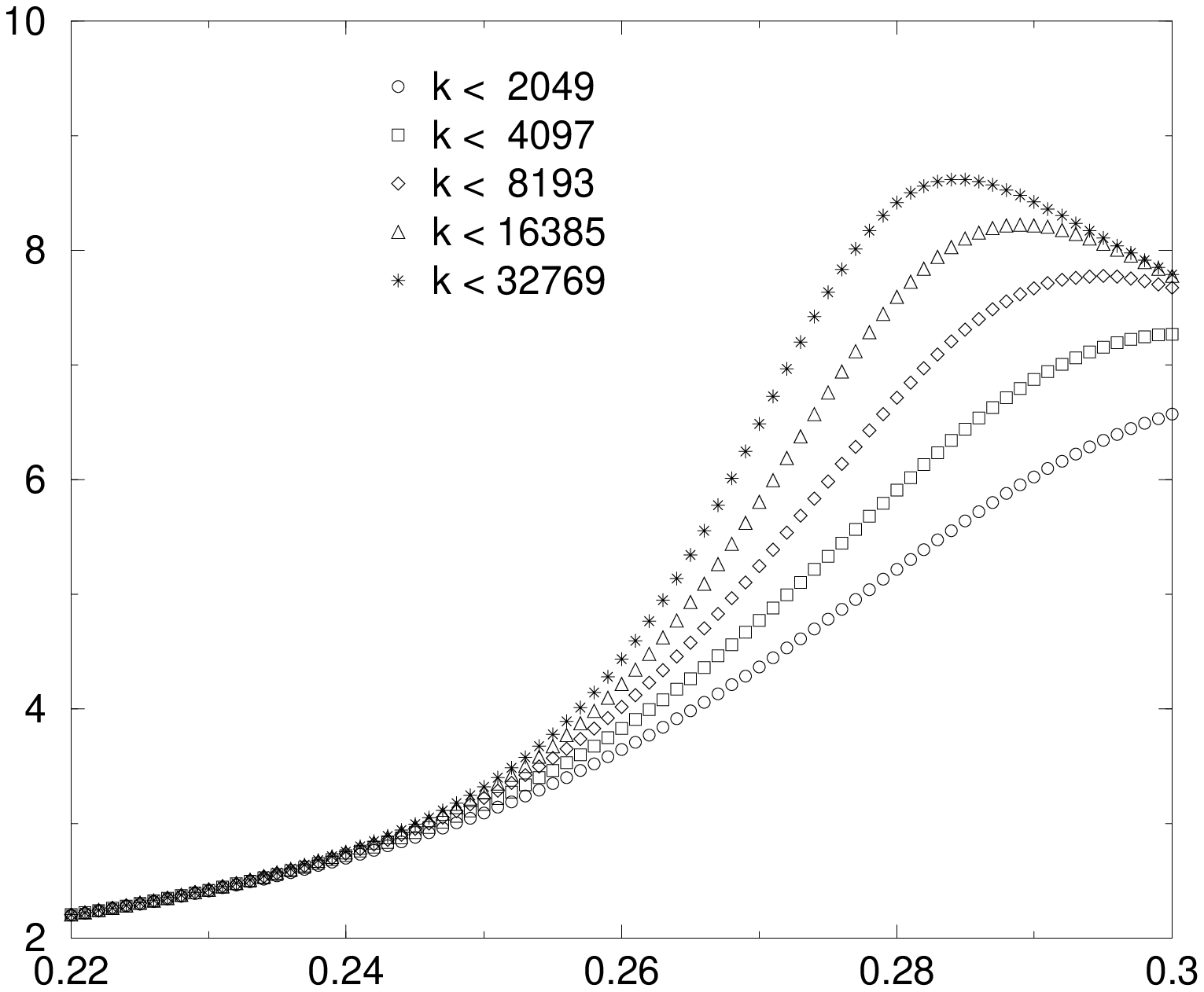}
\end{minipage}     
    \caption{\em Partial resummations of $P_k$ and $kP_k$. Left: The
      fraction of sites occupied by the giant component. Right: The
      variation of $\sum kP_k$ close to the transition.}
    \label{fig:perconum}
  \end{center}
\end{figure}

To discriminate between the two situations, $P_\infty$ may be computed
numerically by evaluating a large number of coefficients $P_k$ using
the recursion relation (\ref{Crecur}). As we shall see later, the
convergence of the series is slow for the whole relevant range of
$\mona$. The result of such a partial summation $ \sum_{k
  \leq k_{max}} P_k$ is plotted in 
Fig.\ref{fig:perconum} for $k_{max}=2^{11},\cdots,2^{15}$. It reveals
a phase transition at a value $\mona _c$ between $.24$ and $.29$, going
from a regime where finite components contain all vertices to a regime
where they do not. Below $.24$ and above $.29$, the plots
corresponding to different values of $k_{max}$ are hard to
distinguish, but in the transition region, large values of $k$ make
substantial contributions to $ \sum P_k$. The transition is also
manifest on an analogous study of $ \sum kP_k$.  We shall show later
that $1/4$ is the exact threshold in this model and that $ \sum kP_k$
is discontinuous at the transition. 

The growth of $P_\infty$ just above the threshold seems to start with
many vanishing derivatives, in strong contrast with what happens in
the Erd\"os-Renyi random graph, for which the growth of the giant
component is linear close to the transition. This can be related to
the following observations:

-- As we have recalled in Appendix A, in the Erd\"os-Renyi model the components
of size $k$ occupy a fraction
$\frac{k^{k-1}}{k!}\mona^{k-1}e^{-k\mona}$. As a function of $\mona$,
this fraction has a single maximum at $\mona=\frac{k-1}{k}$. These
values accumulate at $\mona=1^-$, the well-known transition point for
the standard random graph. Then for $\mona \geq 1$, the fraction
of sites occupied by components of size $k$ decreases with a finite
slope  for all $k$'s, and so does the sum, so that the growth of the
giant component is linear close to the transition. 

-- In the model studied in this paper, the behavior of $P_k(\mona)$ as a
function of $\mona$ for generic $k$ is not so easy to get at. However,
a simple numerical analysis leads to the following picture:
$P_1$ is a decreasing function of $\mona$, but for $k >1$, $P_k$ has a
single maximum, at say $\mona_k$. This sequence starts with $\mona_2
\simeq .241$, $\mona_3 \simeq .311$ $\mona_4 \simeq .341$, is maximum
for $k=12$ with $\mona_{12} \simeq .375$ and then decreases very slowly
($\mona_{100} \simeq .338$, $\mona_{1000} \simeq .301$, $\mona_{10000}
\simeq .282$), apparently getting closer and closer to $1/4$.
At the transition, most
$P_k(\mona)$'s are still increasing, and close enough to the transition a
finite but large number of them is still increasing. So subtle
compensation mechanisms can take place, leading possibly to the
vanishing of (infinitely) many derivatives of $P_\infty$ at
$\mona=1/4$.

This is confirmed in the following
subsections, which are also devoted to a more precise description of
the distributions of finite and infinite clusters.  To summarize:
\begin{eqnarray}
P_\infty &=& 0,\quad {\rm if}\quad  \mona \leq 1/4,\nonumber\\
P_\infty &>& 0,\quad {\rm if}\quad  \mona > 1/4.\nonumber
\end{eqnarray}

\section{Behavior below the transition, $\mona <\mona _c$.}

We turn to the examination of the consequences of eq.(\ref{Cdiff}),
the equation that determines the generating function for the number of
clusters of given finite size.

We know that $R(\mona)$, the radius of convergence of $C(z)$, is at
least $1$. To analyze the behavior of $C(z)$ around $z=1$, we define
\begin{eqnarray}
F(\tau)\equiv \mona  -\mont - \mona \partial_{\mont} 
Y (\mont), \quad Y(\mont)= C(e^{\mont}).
\label{Fdef}
\end{eqnarray}
From eq.(\ref{Cdiff}) $F$ satisfies
\begin{eqnarray}
\mona \, (1-e^{-F})\, \partial_{\mont} F = -F - \mont
\label{Feq}
\end{eqnarray}

Below the transition, there is no percolating cluster so that 
$\partial_{\mont} Y(0)=\sum_k kC_k=1$ or alternatively,
\begin{eqnarray}
\sum_{k\geq 1} P_k=1, \quad \mona <1/4 
\label{mesP}
\end{eqnarray}
This makes clear that the normalized positive numbers $P_k$ are the
probabilities for a vertex to be in a connected component of size $k$.
For $F(\mont)$ this translates into the boundary condition
$$
F(0) =0,\quad \mona <1/4
$$

\subsection{Scaling laws.}
\label{sec:scal}

We first look for a formal solution $F_{tr}(\mont)$ of eq.(\ref{Feq})
in the form of a Taylor series in $\mont$:
\begin{eqnarray}
F_{tr}(\mont)= -\mont-\mona  \sum_{m\geq 1}\, 
\frac{\mu_m}{m!}\,\mont^{m} \label{trunc}
\end{eqnarray}
The Taylor coefficients are the moments of the measure $P_k$:
$\mu_m=\sum_k k^m\,P_k$. For example $\mu_1$ is the average proportion
of vertices per component. As eq.(\ref{Cd2}) the differential equation
eq.(\ref{Cdiff}) may be turned into a second order differential
equation for $Y(\mont)$,
\begin{eqnarray}
\partial{_\mont} Y+ \mona \partial^2{_\mont} Y= (Y+\mona
\partial{_\mont} Y)\,(1+\mona \partial{_\mont}^2Y) \label{Ydiff} 
\end{eqnarray}
Eq.(\ref{Cdiff}) with $\partial{_\mont} Y(0)=1$ gives $Y(0)=1-\mona $. 
Eq.(\ref{Ydiff}) then allows us to recursively compute the $\mu_m$'s:
\begin{eqnarray}
\mu_1&=& \frac{4}{(1+\infb)^2}= \frac{1 -2\mona  - \sqrt{1-4\mona
    }}{2\mona ^2} \label{mu<}\\ 
\frac{1}{2}(n-1-(n+1)\infb)\mu_n&=& -(1+\mona \mu_1)\mu_{n-1}- 
\mona \sum_{k+l=n+1\atop k,l\geq 2} \left({n\atop
    k}\right)\mu_l(\mu_{k-1}+\mona \mu_k) \nonumber 
\end{eqnarray}
with
\begin{eqnarray}
4\mona =1-\infb^2,\quad b=\sqrt{1-4\mona } \nonumber
\end{eqnarray}
The square root singularity in the expression of $\mu_1$
indicates that the initial boundary condition (\ref{mesP}) becomes
pathological at $\mona =1/4$ and thus signals the percolation transition.
The recursion relation for the higher Taylor coefficients shows that
$\mu_n$ possesses a pole at $\infb=(n+1)/(n-1)$. It actually 
changes sign, from  positive to negative, across the pole.
As a function of $n$, the $\mu_n$'s have a simple pole at $n_*(\infb)$,
\begin{eqnarray}
\mu_n \simeq \frac{{\rm reg.}}{n-n_*(\infb)},\quad n_*(\infb) 
=\frac{1+\infb}{1-\infb}, \quad n \geq 2
\label{simplet}  
\end{eqnarray}
where the numerator does not vanish at $(n-1)=\infb(n+1)$. 
Since $\mu_n=\sum_k k^n P_k$ at least for $n<n_*(\infb)$, this simple pole
codes for the asymptotic behavior of $P_k$ at large $k$.
Indeed, recall that if $P_k\simeq k^{-\nu}$ as $k\to\infty$ then
$\sum_k k^xP_k$ diverges as $x\to \nu-1$ 
from below with a simple pole at this value. 
Thus assuming that we may extend the simple pole (\ref{simplet})
to non-integers value of $n$ we learn that
\begin{eqnarray}
P_k\simeq_{k\to\infty} {\rm const.}\, k^{-\nu(\mona )}
\label{asympk}
\end{eqnarray}
with
\begin{eqnarray}
\nu(\mona )=1+n_*(\infb)=\frac{1+\sqrt{1-4\mona }}{2\mona }
\label{expopk}
\end{eqnarray}
This means in particular that the probability distribution  $P_k$ 
(which describes the proportion of the system occupied by clusters of
size $k$) is large
and only the moments $\mu_n$ for $n<n_*(\infb)$, exist.

The fact that the Taylor coefficients $\mu_n$, computed from the
recursion relation (\ref{mu<}), cannot consistently be interpreted as
moments of a probability distribution for $n>n_*(\infb)$ indicates
that the expansion (\ref{trunc}) is only up to $o(\mont^{n_*})$ terms.
Indeed, the differential equation (\ref{Feq}) is compatible with an
expansion of $F(\mont)$ for $\mont<0$, $z=e^\mont<1$, of the form:
\begin{eqnarray}
F(\mont)=-\mont-\mona\sum_{q,p\geq 0}\,\frac{(-)^q}{q!}\,
y_{p,q}\,(-\mont)^{q+pn_*(\infb)},\quad 
 \nonumber
\end{eqnarray}
with $y_{0,0}=0$ and $y_{0,m}=\mu_m$.  As a function of the complex
variable $z$, $\partial_z C(z)$ has thus a branch cut starting at
$z=1$:
\begin{eqnarray}
\partial_z C(z) \simeq {\rm const.}\, (1-z)^{n_*(\infb)}+\cdots,\quad
{\rm around}\ z=1
\label{cut1}
\end{eqnarray} 
For $n_*(\infb)$ an integer, this formula should becomes
$\partial_z C(z) \simeq {\rm const.}\, (1-z)^{n_*(\infb)} \log (1-z)+\cdots$.
The cut implies the asymptotic behavior 
(\ref{asympk}) for $P_k=\oint_0 \frac{dz}{2i\pi} \partial_z C(z)z^{-k}$.

At the transition but from below, i.e. $\mona =1/4^-$, there are
logarithmic correction to the scaling behavior (\ref{asympk}), and the
branch cut equation is $\partial_z C(z) \simeq {\rm const.}\,
(1-z)/\log (1-z)+\cdots$. This ensures that the first moment is
finite, and its value, computed from (\ref{mu<}), is $\mu_1\vert_{\mona
  =1/4^-}= 4$

\subsection{Scaling domain, $\mona < \mona _c$.}
\label{sec:toto}

The scaling law (\ref{asympk}) may be linked to the typical growth
rate of large clusters in the system. For concreteness, consider the
component of vertex $t'$ for any given $t'$. For very large $t$, the
number of arrows emerging from vertex $t'$ grows like $\mona \log t$.
Then we infer that the size of the genealogical tree of $t'$ will grow
like $t^{\mona}$ (under the hypothesis that the genealogical tree is
indeed tree-like, a reasonable assumption for small $\mona$).  This
counting of descendants gives a crude lower bound for the size of the
connected component of $t'$. Hence we expect that the system contains
components whose sizes grow like a power of $t$. To estimate this
power, we argue as follows.

Consider a given large cluster of size $ k(t) \ll t $ at time $t$:

i) $k(t+1)-k(t)$ is $0$ with probability $q_{t+1}^{k(t)}$ and
$1+\sum_p pn_p(t)$ with probability $1-q_{t+1}^{k(t)} \simeq \mona k(t)/t
$ times the probability that vertex $t+1$ connects to $n_p(t)$ clusters
of size $p$ apart from the large cluster;

ii) removing the given large cluster does not change the thermodynamical
properties of the graph, so the probability that vertex $t+1$ connects
to $n_p(t)$ clusters of size $p$ apart from the large cluster is simply
the probability that vertex $t+1$ connects to $n_p(t)$ clusters of
size $p$. Hence for large $t$, from eq.(\ref{nbis}), $\vev{\sum_p
  pn_p(t)}\simeq \mona \sum_k kP_k=\mona \mu_1$;

iii) suppose we add $\delta t $ new vertices with $\delta t \ll t$ but
$k(t) \delta t \gg t$. Between time $t$ and time $t+\delta t$, many
new clusters have be connected to the given large components so $1 \ll
k(t+\delta t)-k(t) \ll k(t)$, but this has not changed the
thermodynamical properties of the graph. Hence we can average the
equation in i) to get a deterministic equation
\[
k(t+\delta t)-k(t) \simeq  \mona k(t)
(1+\mona \mu_1)\delta t /t.\] 
This leads to
\begin{equation}
\label{eq:grosclu}
 k(t) \sim t^ {1/\nu} \mbox{ with } 1/\nu = \mona(1+\mona \mu_1)=
2\mona /(1+\infb).
\end{equation}  

We find that the growth rate of large clusters is universal. As
expected, their growth exponent is larger than $\mona$, the
genealogical tree growth, because it takes not only descendants
into account but the whole component. The difference is maximum at the
transition, where $\nu=2$. 

The fact that the same exponent, $\nu$, governs the asymptotic 
behavior of $P_k$ at large $k$ and the size of large clusters for 
large $t$ can be understood directly as follows.  First consider
one 
realization of the random graph for a given $t$ and suppose that 
there is a single component of maximal size, say $K$. Observe that 
$\sum_{k \geq l} kN_k(t)$ is by definition the number of points in 
components of size larger than $l$. This number is strictly larger
than $l$ if $l <K$, but vanishes if $l > K$. So $K$ is 
characterized by the identity $\sum_{k \geq K} kN_k(t)=K.$ 
 
From this we infer by taking the average that for large $t$ the relation
\[
\sum_{k \geq k(t)} k\vev{N_k(t)} \approx k(t)
\]
gives a sensible characterization for $k(t)$, the order of magnitude of
the size of large components in the graph.  We write $\sum_{k
  \geq k(t)} k\vev{N_k(t)}=t-\sum_{k < k(t)} k\vev{N_k(t)}$ and use that
for $\mona < 1/4$, $\sum_k P_k=1$ to write $\sum_{k \geq
  k(t)}tP_k+\sum_{k < k(t)} tP_k-k\vev{N_k(t)} \approx k(t)$.
For large $k(t)$, the asymptotics of the first sum is $\sum_{k \geq
  k(t)}tP_k \approx t k(t)^{-\nu+1}$. The second sum is made of finite
size corrections. If we assume that these that these are not too
large, we conclude that  $t k(t)^{-\nu+1}\approx  k(t)$, i.e. that $
k(t)\approx t^{1/\nu}$.

In fact, experience from finite size scaling suggests that the two
sums give contributions of the same order of magnitude. The idea is
that $tP_k$ is the main contribution to $k\vev{N_k(t)}$ not only when
$tP_k$ is of order $t$, but even when simply $tP_k \gg 1$ so that self
averaging remains valid. This means
that $tP_k-k\vev{N_k(t)}$ and $k\vev{N_k(t)}$ become of the same order
of magnitude only when $k$ is so large that $tP_k \approx 1$. So again
we see that $k(t)$ is characterized by $tP_{k(t)} \approx 1$, and 
we conjecture that  $t k^{-\nu}$ is a scaling variable.

\section{Behavior above the transition, $\mona >\mona _c$.}

Above the transition there is an giant component. Let $P_\infty$ be
its relative size. By definition, $\partial{_\mont} Y(0)=1-P_\infty$ 
or alternatively
\begin{eqnarray}
\sum_{k\geq 1} P_k + P_\infty =1,\quad \mona >1/4 
\label{probP}
\end{eqnarray}
This makes clear that the $P_k$'s and $P_\infty$ define the probability
distribution of vertices among the clusters of different sizes, 
with $P_\infty$ the probability for a vertex to be in the percolating
cluster.

The size of the giant component $k_\infty$ increases linearly with
time~: $k_\infty\simeq t P_\infty$ for 
$t$ large. The slope may be evaluated as follows.
Imagine adding a new vertex at time $t+1$. It is connected to the
percolating cluster with probability $1-q_{t+1}^{k_\infty}$ so that
\[
k_\infty(t+1)\simeq (1-q_{t+1}^{k_\infty})\,(k_\infty(t)+1+\sum_ppn_p(t)) +
q_{t+1}^{k_\infty}\,k_\infty(t) 
\]
where, as in eq.(\ref{Nt}), $n_p(t)$ are the numbers of components of
size $p$ connected to the new vertex. This can be rearranged as
$k_\infty(t+1)-k_\infty(t) \simeq (1-q_{t+1}^{k_\infty(t)})
(1+\sum_ppn_p(t))$. The quantity $P_{\infty}$ is self averaging, so
using $q_t\simeq 1-\mona /t$ and $k_\infty\simeq tP_\infty$ at large
time we infer $P_\infty=(1-e^{-\mona P_\infty})\vev{(1+\sum_ppn_p(t))}$
which by use of eq.(\ref{nbis}) leads to
\begin{equation}
P_\infty = (1-e^{-\mona  P_\infty})\,(1+\mona  \mu_1) 
\label{Pinf1}
\end{equation} 
As shown below the above equation is exact, but it does not determine 
$P_\infty$ as this requires knowing the moment $\mu_1$.

Above the transition, $F(\mont)$ still satisfies the differential equation
(\ref{Feq}) but with a different boundary condition:
\begin{eqnarray}
F(0)= \mona \,P_\infty,\quad \mona>1/4 
\label{FPinfty}
\end{eqnarray}
with $P_\infty$ vanishing at the transition i.e. as $\mona\to 1/4^+$.
This modifies the behavior of its moments. So let us expand
$F(\mont)$ in Taylor series around the origin:
\begin{eqnarray}
F(\mont)=\mona  P_\infty -\mont -\mona \sum_{m\geq 1}\frac{\mu_m}{m!}\mont^m
\nonumber
\end{eqnarray}
with $\mu_m=\sum_k k^mP_k$. Eq.(\ref{Feq}) implies eq.(\ref{Pinf1})
which is thus exact. 
The second order differential equation (\ref{Ydiff}) 
does not fix $P_\infty$ but determines recursively the $\mu_m$'s which depend
parametrically on $P_\infty$.

It remains to decipher what the behavior of the size of the giant
component is, at least close to the transition. This will follow from an
analysis of the behavior of $F(\mont)$ close to its singularities.  As
we are going to show, $F(\mont)$ possesses square root branch cut at a
point $\mont_c>0$ but exponentially close to the origin. This means
that above the transition the singularity of the function $\partial_z C(z)$
is at $z_c=e^{\mont_c}>1$, with $z_c$ depending on $\mona $, a
behavior which has to be compared with the cut at $z=1$ below the
transition, eq.(\ref{cut1}). More precisely, we show that
\begin{equation}
  \label{eq:sing}
  \log P_\infty \sim -\pi/\supb, \qquad \mont_c \sim P_\infty/8e \quad
  \mbox{for } \mona \to 1/4^+
\end{equation}
with $\supb$ defined by 
\[
4\mona \equiv \supb^2+1,\quad \supb \equiv \sqrt{4\mona -1}\ .
\]

\subsection{Preliminaries.}
\label{sec:prel}

To prove eq.(\ref{eq:sing}) we shall look at the behavior of $F$ in the neighborhood
of three different points:

\noindent -- at the origin $\mont=0$ where $F(\mont)$ takes
the boundary value (\ref{FPinfty}), 

\noindent -- at the branch point $\mont_c>0$  with $F(\mont_c)=0$, 
$\partial{_\mont} F(\mont_c)=\infty$ and,

\noindent -- around the point $\mont_d<0$ specified by the condition
$F(\mont_d)+2\mont_d=0$.

\vspace{.3cm}

Let us first show $\mont_c$ and $\mont_d$ exist. The function we are
interested in is defined by $F=-\mont+\mona (1- \sum_{k \geq 1} P_k
e^{k\mont})$, and satisfies the differential equation eq.(\ref{Feq}).
So $\partial_\mont F<0$ and $F$ decreases from $+\infty$ at
$\mont=-\infty$ to $0$ at a point $\mont_c$ which is non-negative
since $F(0)\geq 0$.  $F$ also satisfies the obvious inequalities $0 <
F+\mont < \mona$ for $\mont<0$ and $\partial{_\mont}^2 F <0$. So there
is a single point $\tau _d$, with $-\mona < \mont_d<0$, at which
$F(\mont _d)+2 \mont _d=0$ and $F+2\mont$ has the sign of $\mont -
\mont _d$.  Both $\mont_c$ and $\mont_d$ vanish at the transition,
$\supb\to 0$.  

It turns out that most of these properties could be proved using only the
differential equation. More precisely, take any solution $E$ of
(\ref{Feq}) with an initial condition at  $\mont _i < 0$ such that
$E(\mont _i)+ \mont _i > 0$. The function $E$ can be extended on the right as
a positive function on a maximal interval $[\mont _i, \mont _f[$. Then
$E$ is strictly decreasing for $\mont > \mont _i$, $\lim _{\mont _f^-}
E =0$, $\mont _f \geq
0$, and $\mont _f > 0$ if $\mona > 1/4$. This discussion emphasizes
the intrinsic role played by $\mona = 1/4$ in this problem.

Before proving eq.(\ref{eq:sing}) let us get intuition from a simpler,
but more universal, version of the differential equation (\ref{Feq}).
If we look at a region where $F$ is small  
-- and we know that such regions exist generically --  
we can estimate $1-e^{-F}$ by $F\simeq f$ with $f$ approximating $F$. 
This leads to the simplified, but still non-linear, differential equation 
\begin{equation}
\mona  \,f\, \partial{_\mont} f +\mont + f=0.
\label{eq:univ}
\end{equation}
It turns out that this simpler equation can be solved in closed
form. Before doing that, let us further assume that $\mont$ is close
to $\mont _c$ so that $f$ can be neglected compared to $\mont$.
Then we get $\mona  \,f\, \partial{_\mont} f \simeq - \mont$,
with solution 
\[
f  \simeq \sqrt{(\mont_c^2-\mont^2)/\mona},
\] 
leading to the announced square root singularity. But this
does not give informations on the location of the branch point. 

Consider now the following functional of an arbitrary function $f$ of $\mont$
\begin{equation}
\frac{1}{2}\log(\mona  f^2 + \mont f+\mont^2) - \frac{1}{\supb}
\arctan\left(\frac{\supb f}{f+2 \mont}\right).
\label{eq:int}
\end{equation}
To fix conventions, we specify the function $\arctan$ by demanding
that it is continuous and takes value in $]-\pi/2,\pi/2[$. The total
derivative of this functional with respect to $\mont$
is 
\[
\frac{\mona   \, f \, \partial{_\mont}f + \mont +f }{\mona  f^2 +
  \mont f+\mont^2}.
\]
It vanishes if $f$ is a solution of eq.(\ref{eq:univ}),
so we have indeed solved in closed form equation (\ref{Feq})
for small $F$. As this is the domain we are interested in, it
is tempting to argue that $F$ and $f$ should exhibit the same singular
behaviour. This turns out to be true, but there are some subtleties
because the limit $\mona \to 1/4^+$ is singular: the size of the
domain for which $f$ is a uniform approximation to $F$ shrinks to
$0$.

\subsection{Rigourous estimates}
\label{sec:rigest}

Our strategy is to use the invariant (\ref{eq:int}) for
the approximate equation (\ref{eq:univ}) to derive exact inequalities for $F$. 

Let us observe that the functional
(\ref{eq:int}) is singular at points where $f+2 \mont=0$ where it has a
jump of amplitude $\pm \pi/\supb$. With this in mind, we
define the functional
\begin{eqnarray}
  \label{eq:Fint}
  I(\mona ', F(\mont),\mont) 
& \equiv & \frac{1}{2}\log(\mona '  F^2 + \mont F+\mont^2) \\
& & -
\left\{ 
    \begin{array}{lcr}
\frac{1}{\supb'} \arctan\left(\frac{\supb ' F}{F+2 \mont}\right)+
\frac{\pi}{\supb'}& \rm{ 
  if } & \mont  < \mont _d<0 \\ \frac{\pi}{2\supb'} & \rm{ if } &
\mont= \mont _d \\ 
\frac{1}{\supb'} \arctan\left(\frac{\supb' F}{F+2 \mont}\right)
& \rm{if } & \mont _d <  \mont <   \mont _c 
    \end{array}
\right. \nonumber
\end{eqnarray}
In this definition, $\mona ' =(1+\supb'^2 )/4> 1/4$ is a priori
independent of $\mona$, the value for which $F(\mont)$ is considered.
The functional $I(\mona ', F(\mont),\mont)$ is a smooth function of
$\mont$ on $]-\infty,\mont _c[$. Using the differential equation for
$F$, its total derivative with respect to $\mont$ is found
to be
\begin{equation}
  \label{eq:deriv}
 \frac{dI }{d\mont}= \frac{\mont+F}{\mona '  F^2 + \mont F+\mont^2}\left(
1-\frac{\mona '}{\mona}\frac{F}{1-e^{-F}}\right).
\end{equation}

If $\mona '=\mona$, the right-hand side is always negative, so $I$ is
decreasing, and comparing its values at $\mont<\mont _d$, at $\mont
_d$, at $0$ and at $\mont _c$ we find 
\begin{equation}\label{eq:bornsup}
I(\mona, F(\mont),\mont)  > \log (\supb|\mont
_d|e^{-\pi/2\supb}) > \log (\mona ^{3/2}
P_\infty)-\frac{1}{\supb} \arctan\supb >\log \mont _c. 
\end{equation}
We know that $\mont _d > -\mona$, so taking a fixed $\mont < -1/4$, we
can take the limit $\mona \to 1/4^+$. At point $\mont$, $F$ is
analytic in $\alpha$, so $I(\mona, F(\mont),\mont)+\pi/\supb$ has a finite limit,
and we get exponentially small upper-bounds for $|\mont _d|, P_\infty$
and $\mont _c$. For instance, $|\mont _d| \supb e^{\pi/2\supb}$ is bounded
above when $\mona \to 1/4^+$.

For $\mona '< \mona$,  $\frac{dI }{d\mont}$ changes sign at a
point $\mont'$ solution of $1-\frac{\mona'}{\mona}\frac{F}{1-e^{-F}}=0$.
This point is unique as $F$ is decreasing and goes from $+\infty$ at
$-\infty$ to $0$ at $\mont _c$. Let us choose $\alpha'$ such that
$\mont'\leq\tau_d$ -- and such choices exist. Then 
$\frac{dI }{d\mont}$ goes from negative to positive if $\mont$
increases so that $I$ increases for $\mont$ varying from
$\mont'$ to $\mont_c$. This leads to lower bounds
\begin{equation} \label{eq:borninf}
I(\mona',F(\mont'),\mont')<
\log (\supb'|\mont _d|e^{-\pi/2\supb'}) <  \log ({\mona'} ^{1/2}
\mona P_\infty)-\frac{1}{\supb'} \arctan\supb' < \log \mont_c.
\end{equation} 

Take $\mona'/\mona=(e^{2\mont_d}-1)/2\mont_d$,
in such a way that $\frac{dI}{d\mont}$ vanishes exactly at $\mont_d$. 
Then $\mona ' =\mona(1+\mont _d+\cdots)$, and $\supb ' =\supb (1+\mont
_d/2\supb^2+\cdots)$ when $\mona \to 1/4^+$. Comparing the lower
bounds with the upper bounds obtained above, we see that
\[ 
\supb |\mont _d| e^{-\pi/2\supb}\sim P_\infty/8e \sim \mont _c \quad
\rm{for } \quad \mona \to 1/4^+.
\]

To get a lower bound for $|\mont _d|$ we take $\mona'$ such that
$\frac{dI}{d\mont}$ vanishes at $\mont' < \mont _d$.  Then
$-\frac{1}{\supb'} \arctan\left(\frac{\supb 'F}{F+2 \mont}\right) >0 $
and $\mona' F^2 + \mont F+\mont^2 = {\supb'} ^2 F^2/4 +(\mont+F/2)^2
\geq {\supb'} ^2 F^2/4$ so we have a crude bound
$I(\mona',F(\mont'),\mont')+\pi/\supb' >\log\supb' F(\mont')/2$.  Taking for
instance $\supb'=\supb-\supb^{2}$, so that $\supb'/\supb\to 1$ and
$F(\mont')\to 0$ as $\supb\to 0^+$.  Reporting in
$\frac{dI}{d\mont}=0$ yields $F\sim 4\supb^{3} $. Thus, from
(\ref{eq:borninf}), $|\mont _d| \supb^{-3}e^{\pi/2\supb}$ is bounded
below when $\mona \to 1/4^+$.

The above upper and lower bounds then imply
\[
  2\log\mont_d \sim \log \mont_c \sim \log P_{\infty}  \sim-\pi/\supb
  \quad \mbox{as} \; \supb\to 0^+. 
\]

We restate the physically most important result: up to an algebraic
prefactor, the size of the percolation cluster close to percolation is
\begin{equation}
  \label{eq:pof}
  P_\infty \propto e^{-\pi/\sqrt{4\mona-1}}.
\end{equation}

In fact, we have obtained a better estimate. We expect that 
$\mont_d \supb^{1-\gamma}e^{\pi/2\supb}$ as a finite limit for $\supb
\to 0^+$ for a certain $\gamma$, or equivalently
\begin{equation}
  \label{eq:equiv}
  P_\infty \sim \rm{const}\; \supb^{\gamma}e^{-\pi/2\supb} , \quad \supb
\to 0^+.
\end{equation} 
We have proved that if $\gamma$ exists, $0\leq \gamma \leq 4$. 
The
asymptotic behavior of the probabilities $P_k$ is not the same below
and above the transition as $C(z)$ does not have the same analytical
properties on the two sides of the transition. 

Since above the
transition the branch point
is located at $z_c=\exp \mont_c>1$, the $P_k$'s now decrease
exponentially. More precisely, $\partial_z C(z)$ possesses a square
root branch point at $z_c$,
$$
\partial_z C(z)= {\rm const.}\, \sqrt{z_c-z} + \cdots
$$
so that $P_k=\oint \frac{dz}{2i\pi} z^{-k} \partial_z C(z)$ behave as
\begin{eqnarray}
P_k \simeq_{k\to\infty} {\rm const.}\  k^{-3/2}\, z_c^{-k}
\simeq {\rm const.}\  k^{-3/2}\, e^{-k\, \mont_c},
\label{Pklarge}
\end{eqnarray}
to be compared with eq.(\ref{asympk}). 
\subsection{Universality.}
\label{sec:univ}

We give now a universality argument suggesting that $\gamma=0$ and this is
confirmed by solving numerically  the differential equation
(\ref{Feq}). 

Our invocation of universality rests on eq.(\ref{Pklarge}), in which   
$\mont_c$, or equivalently (in the vicinity of the critical point)   
$P_\infty/8e$, controls the exponential decrease of the $P_k$'s and
plays the role of a mass gap. This mass gap is
exponentially small close to the transition, and by analogy we may
argue that increasing $\supb$ is a marginally relevant perturbation of
the percolating critical point. Introducing the Wilson-Callan-Symanzik
beta function $B(\supb)$, we expect a relation
\begin{equation}
  \label{eq:Beta}
\mont_c \simeq P_\infty/8e  \simeq \exp \int ^\supb \frac{d\,\supb'}{B(\supb')}.
\end{equation}
Comparison with our formula gives $B(\supb)=\supb^2/\pi+B_3\supb^3+\cdots$.
It is known from field theory that the coefficient $B_3$, which
dictates the exponent of the algebraic prefactor, is universal.

Such universal features are controlled by the continuum limit. In our
framework, the continuum region is reached at small $\mont$, and small
$F$, and we expect that the continuum limit is governed by
(\ref{eq:univ}). Note that both $\mont_c$ and $|\mont_d|$ are
exponentially small close to the transition, and $F(\mont)$ remains
small for $\mont$ between these points.  This a posteriori justifies
looking at the approximate equation (\ref{eq:univ}). Consequently, as
far as universal quantities are concerned, the inequalities in
(\ref{eq:bornsup}) can be replaced by equalities.  This leads to
$\gamma=0$, or $B_3=0$ i.e.
\[
 P_\infty \sim \rm{const}\; e^{-\pi/2\supb} 
\]
as announced. 

As universality could suggest, eq(\ref{eq:univ}) turns out to
describe the continuum limit for a larger class of evolving networks
than just the specific one we are studying. This is the case for the
model studied in \cite{new}. We refer to the original paper for the
definitions. It suffices to say that eq.(\ref{Feq}) is replaced by
\[
2\delta S \partial_{\mont}S=-S-(e^{\mont}-1).
\]
In this equation, $1-S$ is a generating function, the coefficient of
$e^{k\mont}$ giving the fraction of points in components of finite
$k$, so $S(0)$ is the fraction of sites occupied by the giant
component.  The parameter $\delta$ is the average number of edges
created at each time step, so this is is precisely the equivalent of
our $\mona$. To study $S$ close to $\mont=0$, the approximation is
$e^{\mont}-1 \sim \mont$, and we retrieve (\ref{eq:univ}), with
$\mona$ replaced by $2\delta$. The percolation threshold is
$\delta=1/8$ (as expected, percolation thresholds are not universal),
and the size of the infinite component behaves like
$e^{-\pi/2\sqrt{2}(\delta-1/8)^{1/2}}$. The 
prefactor $\pi/2\sqrt{2} \simeq 1.111$ compares quite well with the
numerical value $1.132 \pm 0.008$ obtained in the original paper.
This fact was already noted in \cite{doro2}.

\subsection{Scaling regime, $\mona >\mona _c$.}

We have accumulated evidence that the percolation phase
transition is very similar to
the Kosterlitz-Thouless phase transition in the XY model. For
instance, the probabilities $P_k$ follow the scaling laws
(\ref{asympk}) below the transition with scaling exponents varying
continuously with the parameter $\mona $ while they decrease
exponentially above the transition. 

The discussion of universality in the previous section suggests to
look for scaling functions describing the neighborhood of the critical
point. For instance, from eq.(\ref{Ydiff}) it follows that the moments
$\mu_m$, $m\geq 2$, diverge as $P_\infty\to 0$ with
\begin{eqnarray}
\mu_m \simeq_{\mona\to1/4^+} \frac{g_m}{P_\infty^{m-1}}, \quad g_1=12,\
g_2=64, \cdots \nonumber
\end{eqnarray}
In particular, $\mu_1$ is discontinuous at the transition,
jumping from $4$ to $12$. These scaling relations imply that the
function  
$$
G(x)\equiv P_\infty^{-1} F(xP_\infty)
$$
has a finite limit, denoted by $G_c(x)$, as $\mona \to 1/4^+$ for
any fixed $x$,
\begin{eqnarray}
G_c(x)=\mona -x-\mona \sum_{m\geq 1}\frac{g_m}{m!}x^m
\label{defgc}
\end{eqnarray}
The differential equation (\ref{Feq}) at $\mona =1/4^+$ then reduces
to $G_c\partial_xG_c+4(G_c+x)=0$. With the boundary condition
$G_c(0)=1/4$, this is integrated as:
\begin{eqnarray}
\log\left(4(G_c+2x)\right) + \frac{2x}{G_c+2x}=0 \label{eqgc}
\end{eqnarray} 
The left-hand side is very reminiscent of the limit
$\mona \to 1/4$ of eq.(\ref{eq:int}). 
As a consequence $G_c(x)$ has a square root branch point at $x_c=1/8e$
at which $G_c$ vanishes and $g_m/m!\simeq m^{-3/2}(8e)^m$ for $m$ large.
Actually, a more precise computation based on the Lagrange formula
presented in Appendix C  yields the exact value of the scaling
coefficients $g_m$:
\begin{eqnarray}
g_{m+1} = m^m\, 8^{m+1},\quad m\geq 1. \label{gscal}
\end{eqnarray}

\section{Chronological profiles}
\label{sec:chropro}
One can make a direct counting of the average
number of connected components in the random graph at time $t$ which
are copies of a given finite labeled graph. This will allow us to
retrieve the results of section \ref{sec:clusters}. But this also
leads to a detailed local-in-time description of the connected
components which illustrates the consequences of the chronological
memory of our model. 

\subsection{Tree distributions.}

Let $G$ be a labeled graph with vertices $1,2, \cdots,k$. We let $m_j$ be the
number of edges connecting vertex $j$ to a vertex with smaller label,
so that $m=m_1+\cdots+m_k$ is the number of edges of $G$. 
We look for the average number of increasing maps $v$ from
$[1,\cdots,k]$ to $[1,\cdots ,t]$ such that the vertices $v_1,\cdots,
v_k$ span a connected component of the random graph isomorphic to $G$.
This number is the average number of connected components isomorphic 
to $G$ in the random graph.

By the rules of construction of the random graph, the probability
that the vertices $v_1,\cdots, v_k$ with $1\leq v_1 < \cdots < v_k
\cdots \leq t$ span a connected component of the random graph
isomorphic to $G$ is
\[ 
\prod_{i=1}^k \left((1-q_{v_i})^{m_i} q_{v_i}^{v_i-1-m_i}\prod_{v_i
    < w_i < v_{i+1}} q_{w_i}^i \right)
\]
with the convention $v_{k+1}\equiv t+1$.

The average we look for is obtained by summing this expression over
the $v_i$'s. For large $t$, using the asymptotic behavior $p_j \sim
\mona /j $ for large $j$, the sum
can be formally reinterpreted as a Riemann sum, leading to a
contribution
\[
 t^{k-m}e^{-k\mona }\mona ^m\int_{0\leq \sigma_1 \leq \cdots \leq
  \sigma_k\leq 1} d\, \sigma_1\, \sigma_1^{\mona -m_1}\cdots d\,
\sigma_k\, \sigma_k^{\mona -m_k}.
\]
In deriving this formula, we have not treated carefully
the contribution of small values of the  $v_i$'s. This is reflected in
the fact that the integral can be divergent if some $m_i$'s are too
large. However, the prefactor $t^{k-m}$ is the sign that in this case the
sum over the $v_i$'s is nevertheless negligible in the large $t$
limit, as a more careful treatment would show. 

The most salient feature of this formula is that only connected graphs
with $k=m+1$ give a contribution proportional to $t$, i.e.
contribute to $C_k$. Since $k$ is the number of vertices and $m$ the
number of links, this relation characterizes trees as follows from the Euler
formula: only trees contribute to the thermodynamic limit. A given
tree on $k$ vertices with incoming degrees $m_i$, $i=1,\cdots,k$ gives
a contribution
\begin{equation} \label{eq:tree}
\frac{e^{-k\mona }\mona ^{k-1}}{
(\mona +1-m_1)(2\mona +2-m_1-m_2)\cdots
(k\mona +k-m_1-\cdots-m_k)}\end{equation}
 to $C_k$. Observe that for a
tree, $m_1+\cdots+m_i \leq i-1$ for $i=1,\cdots,k$ so that all
integrals are well-defined and finite for real non-negative $\mona $.
It is amusing to note also that the contribution of a single tree of
size $k$ can contain poles in $\mona $ at values that are not in the
list $-1,-1/2,\cdots,-1/k$. These poles have to cancel between
different trees in the sum over trees of size $k$, because we know
that they are absent in $C_k$. But we have no simple explanation for
this cancellation.

This explicit formula makes it easy to show that $C(z)$ is (complex)
analytic in $\mona$ in a neighborhood of $]0,+\infty[$ for every $z$
such that $|z|<1$. Indeed, we know that for $\mona \in ]0,+\infty[$,
and $|z|<1$ the series $\sum_k P_k(\mona)z^k$ is absolutely
convergent. But $P_k(\mona)$ is a sum of non-negative contributions,
each tree giving a contribution of the form (\ref{eq:tree}). Now
suppose that $\mona$ is complex with positive real part. For each
tree contribution and for fixed $\Re \mona$, the modulus of
\begin{eqnarray*} 
\frac{e^{-k\mona }}{
(\mona +1-m_1)(2\mona +2-m_1-m_2)\cdots
(k\mona +k-m_1-\cdots-m_k)}
\end{eqnarray*} 
is maximal when $\Im \mona=0$ and then the expression is 
real and positive. This is because the
statement is true for every factor. Taking the sum over trees we infer
that $P_k(\mona) \leq P_k(\Re \mona)\left(\frac{|\mona|}{\Re
\mona}\right)^{k-1}$. So the series $\sum_k P_k(\mona)z^k$ is
absolutely convergent if $|z\mona|<\Re \mona$. This equation defines,
for fixed $|z|<1$, a neighborhood of $\mona \in ]0,+\infty[$ in which
we have an absolutely convergent sum of analytic functions analytic of
$\mona$. Hence the sum is analytic in $\mona$ as claimed.

To resum more explicitly the contribution
of all trees of a given size, we need the generating function for labeled
trees with given incoming degrees. Suppose more generally that we give
a weight $x_i$ for each edge leaving vertex $i$ (i.e. connecting $i$
to a $j>i$) and a weight $y_i$ for each edge entering vertex $i$ (i.e.
connecting $i$ to a $j<i$). The generating function ${\mathcal T}$ for
weighted trees on $n$ vertices factorizes nicely as
\begin{eqnarray*}
{\mathcal T} & = & x_1 (y_2x_1+y_2x_2+y_3x_2+\cdots+y_kx_2) \\
& & ~~ (y_3x_1+y_3x_2+y_3x_3+y_4x_3+\cdots+ y_nx_3)\cdots \\
& & ~~ (y_{n-1}x_1+\cdots+y_{n-1}x_{n-1}+y_nx_{n-1})y_n.
\end{eqnarray*}
This generalization of the famous Caley tree formula\footnote{Which
  states that there are $n^{n-2}$ labeled trees on $n$ vertices.}
implies it immediately. It seems to be little known, although it is
implicit in the mathematical literature \cite{bergeron,goulden}. Gilles
Schaeffer \cite{schaeffer} provided us with a clean proof using a refined
version of one of the standard proofs of the Caley tree formula,
putting trees on $k$ vertices in one to one correspondence with
applications from $[1,\cdots,k]$ to $[1,\cdots,k]$ fixing $1$ and $k$,
see e.g. \cite{vanlintwilson}.

This formula can be specialized to $x_i=1$ and
$y_i=1/\sigma_i$ for $i=1,\cdots,k$ to give
\begin{eqnarray} 
C_k&=&e^{-k\mona }\mona ^{k-1} \int_{0\leq \sigma_1 \leq \cdots \leq
  \sigma_k\leq 1} d\, \sigma_1 \cdots d \, \sigma_k\,
(\sigma_1\cdots\sigma_k)^{\mona } \label{Cloctime}\\
&& (\frac{2}{\sigma_2}+\frac{1}{\sigma_3}+\cdots
+\frac{1}{\sigma_k})(\frac{3}{\sigma_3}+\frac{1}{\sigma_4}+\cdots
+\frac{1}{\sigma_k})\cdots(\frac{k-1}{\sigma_{k-1}}+\frac{1}{\sigma_k})
\frac{1}{\sigma_k}.\nonumber
\end{eqnarray}

If one integrates only over a subset of the $\sigma$'s, one gets
marginal distributions. For instance, for $k=1$, we get that in the
thermodynamic limit the fraction of sites with age close to $t\sigma$
that are isolated is $e^{-\mona} \sigma^\mona$. For $k=2$, if we
integrate over $\sigma_2$, we get that the fraction of sites with age
close to $t\sigma$ that are the older vertex of a tree on two
vertices is $e^{-2\mona} (\sigma^\mona-\sigma^{2\mona})$ while if we
integrate over $\sigma_1$, we get that the fraction of sites with age
close to $t\sigma$ that are the younger vertex of a tree on two vertices
is $e^{-2\mona}\sigma^{2\mona}\frac{\mona}{\mona+1}$. The sum,
$e^{-2\mona}(\sigma^\mona-\sigma^{2\mona}\frac{1}{\mona+1})$, gives
the probability that a site with age close to $t\sigma$ belongs to a
tree on $2$ vertices.

Our explicit representation in terms of trees shows that in this
model, and at least for questions concerning connected components, the
thermodynamic limit applies not only to the full system, but also to
slices of fixed relative age $\sigma=t'/t$. In the next section we
shall study $\sigma$-dependent profiles. 

For small $k$'s, we have done  all the integrals and checked the
agreement with the value of $C_k$ obtained by the recursion relation.
But a general proof valid for all $k$'s is lacking.

\subsection{Dating finite components.}

To illustrate the evolving nature of our model, 
we now determine the local-in-time distribution of the cluster sizes.
This means determining for any given age interval what is the proportion
of vertices of these ages which belong to clusters of given size.

Define $p_k(t,t')$ to be the
probability that vertex $t'$ belongs to a component of size $k$ at
time $t$. Guided by the previous tree representation, we infer
that in the thermodynamical limit 
$\sum_{t' \in [t\sigma,t(\sigma+d\sigma)]}\,  p_k(t,t')\simeq
t\rho_k(\sigma)d\sigma$, with $\rho_k(\sigma)$ a deterministic function.
By construction, $\int_0^1 \rho_k(\sigma)d\sigma=kC_k$, the total
fraction of points that belong to components of size $k$. 

The reasoning leading to (\ref{Cdiff}) can be generalized: one writes
down a recursion relation for $p_k(t+1,t')$ and then takes the
average, a step justified by the explicit tree representation. The
event that vertex $t'$ belongs to a component of size $k$ at time
$t+1$ is the exclusive union of several events.

i)  Vertex $t'$ belonged to a component of size $k$ at
time $t$ and this component is not linked to the new vertex $t+1$.
This has probability $p_k(t,t')q_{t+1}^k$.

ii)  Vertex $t'$ belonged to a component of size $l<k$ at time $t$, this
component is linked to the new vertex $t+1$, and together with the other
components linked to $t+1$ (say $n_m(t)$ components of size $m$), it builds a
component of size $k=l+1+\sum_m mn_m(t)$. This has probability
\[ 
p_l(t,t')(1-q_{t+1}^l) \prod_m \bin{N_m(t)-\delta_{m,l}}{n_m(t)}
q_{t+1}^{m(N_m(t)-\delta_{m,l}-n_m(t))}(1-q_{t+1}^m)^{n_m(t)}
 \]

\noindent To perform the explicit sum over $l$ and the 
$n_m(t)$'s, we introduce again generating functions and set
$p_{t,t'}(z)=\sum_l p_l(t,t')z^l$. This leads to
\begin{eqnarray*}
p_{t+1,t'}(z) & = & p_{t,t'}(zq_{t+1}) \\ & & \hspace{-.7cm}
+\left(\sum_l p_l(t,t') 
\frac{z^l(1-q_{t+1}^l)}{q_{t+1}^l+z^l(1-q_{t+1}^l)}\right)
\prod_m (q_{t+1}^m+z^m(1-q_{t+1}^m))^{N_m(t)}
\end{eqnarray*}

In the large $t$ limit this complicated formula simplifies if we use
again the hypothesis of self-averaging and asymptotic independence.
Defining 
\begin{eqnarray}
\rho(\sigma,z)=\sum_k \rho_k(\sigma)z^k,
\label{defrhoz}
\end{eqnarray}
 this leads to 
\[
\sigma\partial_\sigma \rho = \mona (1-ze^{-\mona +\mona z \partial_z
  C})z \partial_z\rho.
\label{profit}
\]

Together with the sum rule relating $\rho$ to $C$, this fixes
completely the profiles $\rho_k(\sigma)$. A relation between
$\rho(1,z)$ and $C(z)$ is obtained by integrating (\ref{profit}) for
$\sigma$ between $0$ and $1$. Using the defining equation for $C(z)$,
eq.(\ref{Cdiff}), this leads to $\rho(1,z)  =  (1+\mona z \partial_z)
C(z)$. Thus, we can summarize our knowledge on component time profiles with
the four relations~:
\begin{eqnarray}
\sigma\partial_\sigma \rho(\sigma,z) & = & \mona (1-\rho(1,z)) z
\partial_z\rho(\sigma,z) \nonumber\\
\int_0^1 d\sigma \, \rho(\sigma,z) & =  & z \partial_z C(z) 
\label{rhoeqss}\\
\rho(1,z) & = & z e^{-\mona +\mona \int_0^1 \rho(\sigma,z)
\,d\sigma} \nonumber\\
\rho(1,z) & = & (1+\mona z \partial_z) C(z). \nonumber
\end{eqnarray}

Expansion in powers of $z$ leads to recursion relations for the
$\rho_k(\sigma)$'s. They can be shown to be alternating
polynomials of degree $k$ in $x=\sigma^{\mona}$ with vanishing
constant coefficient.
The first few polynomials are 
\begin{eqnarray*}
\rho_1 & = & e^{-\mona}x \\
\rho_2 & = & e^{-2\mona}\left(x-\frac{x^2}{\mona+1}\right)\\  
\rho_3 & = & e^{-3\mona}\left(\frac{3\mona+2}{2(\mona+1)}x-
\frac{2}{\mona+1}x^2+\frac{3\mona+2}{2(\mona+1)^2(2\mona+1)}
x^3\right) 
\end{eqnarray*}

Again, we have checked that for small $k$'s the values of $\rho_k$ as
computed from iterated tree integrals and from the generating function
coincide, but we have no general proof.

As an application, let us look at the profile of vertices of relative
age close to $\sigma$ which are the youngest in components of size
$k$. These are vertices that created, when they appeared, a component
of size $k$ -- this happens with probability
$\rho_k(1)=(1+k\mona)C_k$ -- which was then left untouched for the
rest of the evolution -- this happens with probability
$\sigma^{k\mona}$. So the distribution of vertices that are the
youngest in their connected (finite) component is
\begin{equation}
\label{eq:young}
\sum_k (1+k\mona)C_k \sigma^{k\mona} 
=(1+\mona z\partial_z)C(\sigma^{\mona}).
\end{equation}   
In this expression, the formal parameter $z$ has been replaced by
$\sigma^{\mona}$. 

The giant component, when it exists, also has a well-defined profile
to which we turn now.

\subsection{Dating the percolating cluster.}

We now determine the profile of the giant component.

So let $\rho_{\infty}(\sigma)t\, d\sigma$ be the fraction of vertices
whose ages are between $\sigma t$ and $(\sigma +d\sigma\,) t$ which
belong to the giant component. By definition
$\int_0^1d\sigma\,\rho_{\infty}(\sigma)=P_\infty$.  We know from the
tree representation that for finite clusters the thermodynamic limit
applies not only on 
the full graph, but also on time slices. The giant component is the
complement of finite components, so we expect that the density
$\rho_{\infty}(\sigma)$ is self-averaging as is the size of 
the percolating cluster.  To derive an equation fixing this density we
look for the probability for a site of age $j=\sigma \, t$ not to be
in the percolating cluster. On the one hand, by definition of the
density this probability is
$$
{\rm Prob}(j=\sigma\,t \notin [k_\infty]) = 1-\rho_{\infty}(\sigma)
$$
On the other hand, this probability may be evaluated by demanding the
vertex $j$ not to be connected to the older and younger vertices of the
percolating cluster:
$$
{\rm Prob}(j=\sigma\,t \notin [k_\infty])=
\prod_{k<j,k\in[k_\infty]}(1-\frac{\mona}{j})
\prod_{k>j,k\in[k_\infty]}(1-\frac{\mona}{k})
$$
At large time, the first above product converges to
$\exp(-\mona \int^\sigma_0 d\zeta\,\frac{\rho_{\infty}(\zeta)}{\sigma})$ 
and the second to
$\exp(-\mona \int_\sigma^1 d\zeta\,\frac{\rho_{\infty}(\zeta)}{\zeta})$.
So we get the relation:
\begin{eqnarray}
  1-\rho_{\infty}(\sigma) & = &
\exp\left(-\mona \int^\sigma_0 d\zeta\,\frac{\rho_{\infty}(\zeta)}{\sigma}
-\mona \int_\sigma^1 d\zeta\,\frac{\rho_{\infty}(\zeta)}{\zeta}\right)
\nonumber \\ 
 & = & \exp \left(-\mona \int^1_0 d\zeta\, \rho_{\infty}(\zeta)\min
  \left(\frac{1}{\sigma},\frac{1}{\zeta} \right) \right)\label{profil}
\end{eqnarray}

 As $\rho_{\infty}$ is positive, $\int^1_0 d\zeta\, \rho_{\infty}(\zeta)\min
  \left(\frac{1}{\sigma},\frac{1}{\zeta} \right)$ is an increasing
  function of $\sigma$. So $\rho_{\infty}(\sigma)$ is decreasing and has a
  right limit at $0$. Unless $\rho_{\infty} \equiv 0$ (i.e. $\mona \leq
  1/4$), the integral has a logarithmic
  divergence, and henceforth $\rho_{\infty} (0)=1$. More precisely,
\begin{equation} \label{eq:rho_0}
\rho_{\infty} (\sigma)=1-\sigma^{\mona}\exp\left(\mona \int^1_0 d\zeta\,
  \frac{1-\rho_{\infty}(\zeta)}{\zeta}\right) +\cdots 
\end{equation} 
This means that the early vertices belong to the giant component with
probability $1$.  

\begin{figure}[htbp]
  \begin{center}
    \includegraphics[width=0.7\textwidth]{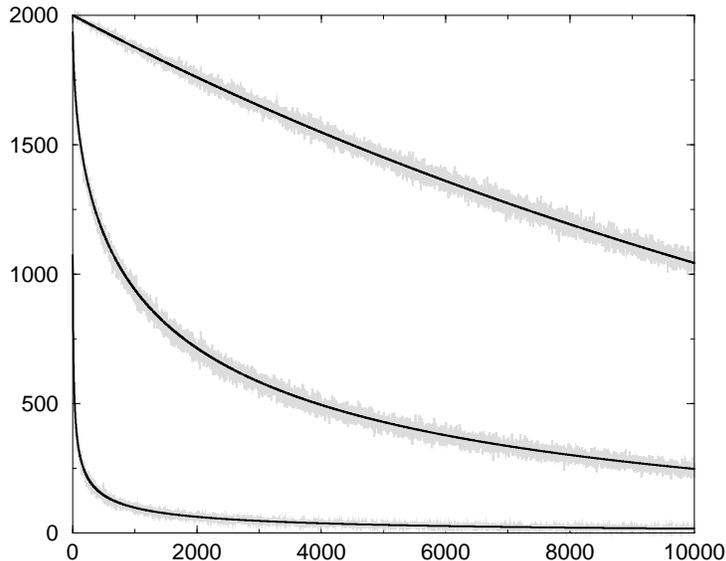}
      \caption{\em The analytic result (solid lines) for the profile
    of the giant component compared to numerical simulations (gray
    clouds) on $2000$ random graphs of size $10000$. From top to bottom, the
    values of  $\mona$ are $1$, $1/2$ and $1/3$.}
      \label{fig:profinfin}
  \end{center}
\end{figure}

On the other hand, by definition  $\int^1_0
d\zeta\,\rho_{\infty}(\zeta)=P_{\infty}$ so taking $\sigma=1$ in
(\ref{profil}) we get that
\begin{equation}
  \label{eq:rho_1}
  \rho_{\infty}(1)=1-e^{-\mona P_{\infty}}.
\end{equation}
This means that the late vertices always belong to the giant component
with a non vanishing probability -- although this probability is
exponentially small close to the threshold. Hence the giant component
invades all time slices above the threshold, and the term
percolation transition is appropriate even with this unusual
interpretation. Our results are illustrated on Fig.\ref{fig:profinfin}.

To conclude this discussion, let us observe that $\rho_{\infty}$ can
be expressed in terms of the profiles of finite components that we
studied before:
\[
\rho_{\infty}(\sigma)=1-\sum_k  \rho_k (\sigma) =1-\rho(\sigma,z=1)
\]
with $\rho(\sigma,z)$ defined in eq.(\ref{defrhoz}).
We end up with yet another equation constraining $\rho(\sigma,z)$,
\[
\rho(\sigma,z=1)=\sigma^{\mona}\exp \left(-\mona +\int^1_0 d\zeta\,
  \rho_{\infty}(\zeta)\min \left(\frac{1}{\sigma},\frac{1}{\zeta}
  \right) \right).
\]

If we expand $\rho(\sigma,z=1)$ in powers of $\sigma^{n\mona}$ as
$\rho(\sigma,z=1) =\sum_{k\geq 1} D_n(1+n\mona)\sigma^{n\mona}$ and
define $D(z)=\sum_{k\geq 1} \frac{D_n}{n}z^n$, we can check that
\begin{equation}
z\partial_zD(z) + \mona  (z\partial_z)^2 D(z) = z \exp(-\mona  -D(z)
+D(1)+z\partial_z D(1))
\label{Ddiff}
\end{equation}
In particular, by differentiation this implies that 
\begin{equation}
(n-1)(1+n\mona )D_n=-\sum_{k+l=n} k^2(1+\mona l)D_kD_l
\label{Drecu}
\end{equation}
The resemblance with equations (\ref{Cdiff}) and (\ref{Crecur}) for
the function $C(z)$ is rather striking, but we have not been able to
use this more deeply. One difference is that (\ref{Cdiff}) determines
$C_1$ from scratch, whereas (\ref{Ddiff}) does it in a two step
process. If $d_n$ is the solution of this equation satisfying $d_1=1$,
then $D_n=d_nD_1^n$, and then $D_1$ has to satisfy
\[
(1+\mona)D_1=\exp(-\mona+\sum_{n\geq 1} \frac{1+n\mona}{n}d_nD_1^n)
\]
Another difference is that the sequence $D_n$ alternates in sign.
Again, the formal parameter $z$ receives a simple physical
interpretation $z=\sigma^{\mona}$ where $\sigma$ labels the relative
date of birth of vertices.

\section{Conclusions}
\label{sec:concl}

In this study, we have described detailed global and local-in-time
features of  evolving random graphs with uniform attachment rules.

Concerning global properties, we have shown that the model has a
percolation phase transition at $\mona=1/4$. Below the transition, the
system contains clusters whose sizes scale like
$t^{(1-\sqrt{1-4\mona})/2}$.  Above the transition, a single
component, the giant component, grows steadily with time. We have
shown that, close to the threshold, the fraction of sites in the giant
component has an essential singularity and behaves as
$e^{-\pi/\sqrt{4\mona-1}}$. The behaviors below and above the
transitions are strongly reminiscent of the two dimensional $XY$
model, so that our model can be interpreted as some algorithmic
equivalent of it, but it seems unlikely that a direct connexion
exists.  This analogy and further scaling properties
we present call for an alternative renormalization group approach to
the transition.

By describing local-in-time profiles, we have shown that they offer a
more accurate vision of the specificities of evolving graphs. It would
be desirable to generalize this approach by answering the following
question : assume that a procedure to assign ages to the vertices of
some evolving graphs has been given, what informations on the
microscopic evolution rules of these graphs can be decoded from the
knowledge of local-in-time statistics~?

\vspace{.5cm}

\noindent Aknowledgements: We thank Gilles Schaeffer for his
clarifying help in tree generating functions and Sergei Dorogovtsev
for his comments and interest.

\section{Appendix A.}

We show how our arguments of section \ref{sec:clusters} can be
modified to describe the Erd\"os-Renyi random graph model. In this
model, one starts with $n$ points, and any two points are connected by
an edge with probability $\mona /n$ (so in this model all the points
are equivalent). Then a limit $n \rightarrow \infty$ is taken. This
famous model describes a static graph, but it can also be rephrased as
an evolving graph in the following way : set $t=n$ and suppose that
points are added one by one, from $1$ to $t$, each new point
connecting to any previous one with probability $\mona /t$. From this
point of view, it can be seen that looking only at the first $t'$
vertices ($t' \leq t$) amount to look at an Erd\"os-Renyi random graph
with a modified connectivity parameter $\mona '=\mona t' / t$. To get
a recursion relation, we start from an Erd\"os-Renyi random graph of
size $t$ with connectivity parameter $\mona$. We add vertex $t+1$ and
connect any older vertex to it with probability $q_{t+1}=\mona/t$, so
that the effective connectivity parameter for the graph on $t+1$
vertices is $\mona (t+1)/t$. Then the derivation proceeds as before,
with the little proviso that in eq.(\ref{Nmoyen}), on the left-hand
side $\vev{N_{t+1}(z)}$ has an effective connectivity parameter $\mona
(t+1)/t$ instead of $\mona$.  So when we take the thermodynamic limit,
nothing changes on the right-hand side of (\ref{Nmoyen}), but an
additional term $\mona \partial_\mona $ contributes to the left-hand
side.  If $c$ denotes the analog of $C$ but for the Erd\"os-Renyi
random graph, then
\[
 \mona \partial_\mona c = -c -\mona  z\partial_z c + z \exp(-\mona  +
\mona  z\partial_z c). 
\]

This little modification in the equation has drastic consequences. The
new equation has a single solution
regular at $\mona=0$, namely
\[
c=\sum_{k\geq 1} \frac{k^{k-2}}{k!}\mona^{k-1}e^{-k\mona}z^k
\]
which is the well-known result. We conclude that our
self-averaging hypothesis 
is valid for the Erd\"os-Renyi model. This makes it more plausible that
it works for our original model as well, a fact also confirmed by
numerical simulations.  

\section{Appendix B.}

Our goal is to prove the following formula 
and to investigate a few of
its consequences. We claim that
\begin{eqnarray}
\vev{\prod_{k\geq 1} w_k^{N_k(t+1)}} = \oint_0 \frac{d\xi}{2i\pi}  
\sum_{j\geq 1}\frac{w_j}{\xi^j} \, 
\vev{\prod_{m\geq 1}[q_{t+1}^mw_m+(1-q_{t+1}^m)\xi^m]^{N_m(t)} }
\label{mast}
\end{eqnarray}
where the contour integral is around the origin.  It is a
Fokker-Planck equation for the Markov process formed by the $N_k(t)$'s,
which could be used to prove systematically that the variables $C_k$
are self-averaging, a task we perform for $C_1$ and $C_2$ at the end
of this appendix.
 
We start from eq.(\ref{Nt}) which may be rewritten as
\begin{eqnarray}
\vev{\prod_k w_k^{N_k(t+1)} } = 
\vev{\prod_j w_j^{N_j(t)-n_j(t)}\, w_j^{\delta_{j;1+\sum_p pn_p(t)}} }
\nonumber
\end{eqnarray}
To compute the last term we insert the tautological identity
\[1=\sum_{k\geq 1}\delta_{k;1+\sum_p pn_p(t)}\]
 in the r.h.s. to get
\begin{eqnarray}
\vev{\prod_k w_k^{N_k(t+1)} } = \sum_{k\geq 1} w_k\,
\vev{\prod_j w_j^{N_j(t)-n_j(t)}\,\delta_{k;1+\sum_p pn_p(t)} } \nonumber
\end{eqnarray}
To compute the r.h.s. expectation value we use a contour integral
representation of the Kronecker symbol
$$
\delta_{k;1+\sum_p pn_p(t)} =\oint_0 \frac{d\xi}{2i\pi} \xi^{-k+\sum_p
  pn_p(t)}
$$
This yields
\begin{eqnarray}
\vev{\prod_k w_k^{N_k(t+1)} } = \sum_{k\geq 1} \oint_0
\frac{d\xi}{2i\pi}\, \left( \frac{w_k}{\xi^k}\right)\,
\vev{ \prod_p (\xi^p/w_p)^{n_p(t)}\, w_p^{N_p(t)}} \nonumber
\end{eqnarray}
The r.h.s. can now be computed using eq.(\ref{ndist}) and gives
eq.(\ref{mast}).

This computation has a rather simple combinatorial
reinterpretation. We set 
\[
F_t(w_1,w_2,\cdots)=\vev{\prod_k w_k^{N_k(t)}},
\]
and observe that going from time $t$ to $t+1$, we add vertex $t+1$ and
edges from the rest of the graph to $t+1$. Suppose that the component
of vertex $t+1$ has size say $k$. This component was build by
``eating'' some components of the graph at time $t$. A component of
size $j$ is swallowed with probability $1-q_{t+1}^j$ and survives with 
probability $q_{t+1}^j$. On the other hand, if one expands 
\[F_t(w_1q_{t+1}+\xi(1-q_{t+1}),w_2q_{t+1}^2+\xi^2(1-q_{t+1}^2),\cdots)\]
in powers of $\xi$, the term of degree $k-1$ enumerates all the
possibilities to build the component of vertex $t+1$ with the correct
probability. Defining $\hat{w}_l(\xi)\equiv
w_lq_{t+1}^l+\xi^l(1-q_{t+1}^l)$,
 summation over $k$ gives 
\[F_{t+1}(w_1,w_2,\cdots)=\sum_k w_k \oint_0 \frac{d\xi}{2i\pi}
\xi^{-k}F_t(\hat{w}_1(\xi),\hat{w}_2(\xi),\cdots),\]
 as obtained previously.

The Fokker-Planck eq.(\ref{mast}) may also be formulated as a
difference equation, similar to a discrete Schrodinger equation.
Let us for instance specify it for $w_k=\sum_j \zeta_j^k$ with
$\zeta_j$, $j=0,\cdots,M$, a set of complex numbers and let
\begin{eqnarray}
\Psi_t(\zeta_0,\cdots,\zeta_M)\equiv \vev{
  \prod_k[\zeta^k_0+\cdots+\zeta_M^k]^{N_k(t)} }
\nonumber
\end{eqnarray}
This parametrization is similar to that used in matrix theory 
where $w_k$ may be thought of as the trace of the $k^{\rm th}$ power of
matrix whose eigenvalues are the $\zeta_j$'s.
The contour integral in eq.(\ref{mast}) can then be explicitly
evaluated by deforming the integration contour to pick the simple pole
contributions located at the points $\xi=\zeta_j$. This gives:
\begin{eqnarray}
\Psi_{t+1}(\zeta_0,\cdots,\zeta_M)= q^t_{t+1}
\sum_j \zeta_j\, \Psi_t(\cdots,\zeta_j/q_{t+1},\cdots) 
\label{diffshro}
\end{eqnarray} 

Eq.(\ref{mast}) or (\ref{diffshro}) may  be used 
to prove that the numbers $C_k$ are self-averaging. 
Let us choose for example two parameters $\zeta_0=1$
and $\zeta_1 = z/t$. Then, only the clusters of size $1$ give a non
trivial contribution to $\Psi_t$ at large time so that
$$
\Psi^{(1)}_t(z)\equiv \Psi_t(1,z/t)
= \vev{\prod_k(1+z^k/t^k)^{tC_k}} \simeq_{t\to\infty} \vev{e^{zC_1}}
$$
At large time, the difference equation (\ref{diffshro}) then reduces
to a differential equation for $\Psi^{(1)}_t(z)$,
$$
(1+\mona )\partial_z \Psi^{(1)}_t(z) = e^{-\mona }\, \Psi^{(1)}_t(z)
$$
It implies that $\log \Psi^{(1)}_t(z)$ is linear in $z$ at large time
which means that $C_1$ is self-averaging and stationary at large time.
More precisely, integrating the above equation gives:
$$
\Psi^{(1)}_t(z)= \exp( z C_1(\mona )),\quad C_1(\mona )= 
\frac{e^{-\mona }}{\mona +1}
$$
in agreement with eq.(\ref{lesC}).

Similarly, to prove that $C_2$ possesses a finite self-averaging limit
as $t\to\infty$ we choose three parameters $\zeta_0=1$,
$\zeta_1=-\zeta_2=\sqrt{z/t}$ so that only clusters of size two survive
in $\Psi_t$ at large time,
$$
\Psi^{(2)}_t(z)\equiv \Psi_t(1,\sqrt{z/t},-\sqrt{z/t})
\simeq_{t\to\infty} \vev{e^{2zC_2}}
$$
Eq.(\ref{diffshro}) then gives a first order differential equation for
$\Psi^{(2)}_t(z)$ which implies that
$$
\Psi^{(2)}_t(z) = \exp(2zC_2(\mona )),\quad
C_2(\mona )=\frac{e^{-2\mona }\mona }{(2\mona +1)(\mona +1)}
$$
in agreement with eq.(\ref{lesC}). Although we do not have a global
argument this proof may clearly be extended to recursively prove
self-averageness of any $C_k$ by choosing the parameters $\zeta_j=\omega^j$
with $\omega^k=z/t$.

\section{Appendix C.}

Here we present the proof of eq.(\ref{gscal}).  Let us first recall
Lagrange formula. Consider a variable $X$ defined by the implicit
relation $f(X)=y$ for some given analytic function $f$.  The solution
of this equation is supposed to be unique so that $X$ is function of
$y$.  Given another analytic function $g(w)$ we look for the Taylor
series in $y$ of $g(X(y))$. This composed function may be presented as
a contour integral:
$$
g(X(y))=\oint \frac{dz}{2i\pi} \frac{f'(z)}{f(z)-y}\, g(z)
$$
Expanding the integrated rational function in Taylor series in $y$
gives Lagrange formula:
$$
g(X(y)) = \sum_{n\geq 0} y^n\, \oint \frac{dz}{2i\pi} 
\frac{f'(z)}{f(z)^{n+1}}\, g(z)
$$

Let $H(x)\equiv 4(G_c(x)+2x)$. Eq.(\ref{eqgc}) translates
into $H(x)\log H(x)=-8x$ or $W(y)e^{W(y)}=y$ for $y=-8x$ and
$H(x)=e^{W(y)}$.
We now apply Lagrange formula with $f(X)=Xe^X$ and $g(w)=e^w$.
This gives
\begin{eqnarray}
H(y)&=& \sum_{n\geq 0}y^n\, \oint \frac{dz}{2i\pi}
\frac{z+1}{z^{n+1}}e^{(1-n)z}  \nonumber\\
&=& 1+\sum_{n\geq 1}\frac{(1-n)^{n-1}}{n!}\,y^n \nonumber
\end{eqnarray}
With $H(x)=4(G_c(x)+2x)$ and $y=-8x$, this proves eq.(\ref{gscal}).

\end{document}